%% file: templateArxiv.tex
\documentclass{article}

\usepackage{PRIMEarxiv}

\usepackage[utf8]{inputenc} 
\usepackage[T1]{fontenc}    
\usepackage{url}            
\usepackage{booktabs}       
\usepackage{amsfonts}       
\usepackage{amsmath}
\usepackage{nicefrac}       
\usepackage{microtype}      
\usepackage{fancyhdr}       
\usepackage{multirow}
\usepackage{graphicx}       
\graphicspath{{media/}}     
\usepackage{hyperref}       
\usepackage[none]{hyphenat}
\usepackage{subcaption}
\title{Resource-Efficient QUBO Formulation for Anchored Currency Arbitrage
\thanks{\textit{\underline{Citation}}: 
\textbf{E. Reinhardt and A. Hauser, Resource-Efficient QUBO Formulation for Anchored Currency Arbitrage.}} 
}

\author{
  Eric A. F. Reinhardt$^1$ and Adam J. Hauser$^1$ \\
  $^1$Department of Physics and Astronomy \\
  University of Alabama, Tuscaloosa, AL 35487, USA \\
  \texttt{eareinhardt@crimson.ua.edu} \\
}
\date{December 2025}

\begin{document}
\maketitle

\begin{abstract}
Currency arbitrage (CA) involves trading currencies in cycles to exploit discrepancies in market valuations. Quadratic unconstrained binary optimization (QUBO) involves minimizing a quadratic cost (energy) function of binary variables. Previous works have explored the use of QUBO to solve CA problems. We build on these previous works by introducing realistic constraints such as beginning cycles from a held currency and accounting for per-transaction trading fees. We show that this formulation requires fewer logical variables (qubits) than previous QUBO encodings in the literature. We derive provably sufficient penalty weights for its constraint terms. We also introduce an exact anchor-gauge reweighting of the exchange rates that compresses the QUBO coefficient range from the rate scale to the arbitrage scale, addressing the finite analog precision of annealing hardware. We demonstrate the efficacy of this formulation using classical simulated annealing against an exact Held--Karp baseline on the same CPU and show that it can effectively find profitable cycles and account for trading fees. Finally, we benchmark faithful implementations of five prior QUBO encodings at matched sampler budgets and show that the proposed encoding is the only one to recover the exact fee-adjusted optimum.
\end{abstract}

\section{Introduction}
Currency arbitrage (CA) is a trading strategy that involves trading based on discrepancies in the valuations of pairs of currencies across markets or exchanges \cite{schrimpf2019,chaboud2023}. In real world applications, CA can involve forecasting of market valuations of different currencies, real-time strategies that involve leveraging latency in market valuations of currencies, or some combination of the two \cite{debelle2011, cartea2019, oomen2017}. Regardless of the source of the imbalance, the task can be reduced to identifying maximally profitable trading cycles in imbalanced markets with the shortest, three-currency case known as ``triangular arbitrage'' \cite{foucault2017, kozhan2012}.

Solutions to the problem of identifying optimal CA cycles can be computationally expensive, with brute-force enumeration scaling as $O(N^3)$ for triangular arbitrage and growing combinatorially, up to $O(N!)$, for unconstrained cycle lengths, where $N$ is the number of currencies being considered. A more typical approach would be to use exponential time algorithms like Held--Karp which scales like $O(2^{N-1}N^2)$ time and $O(2^{N-1}N)$ memory combined with limiting the number of currencies considered \cite{HeldKarp1962}. We also note that merely detecting the existence of an arbitrage cycle can be accomplished in polynomial time via negative-cycle detection on the graph of negated log-exchange rates \cite{bangjensen2000digraphs}; the computationally hard task addressed here is identifying a maximally profitable simple cycle subject to length and anchoring constraints. Due to scaling as well as trading fees, practical CA is generally limited to short cycles, and the empirical literature has focused on the triangular case \cite{foucault2017, kozhan2012}. Prior work has shown that these problems can be represented as quadratic unconstrained binary optimization (QUBO) problems.

In a QUBO representation, the goal is to construct a cost function that is quadratic in binary decision variables and whose global minimum represents the target solution. Many problems can be formatted exactly in this construction \cite{lucas2014}; many more admit QUBO representations with rugged, non-convex energy landscapes on which annealing heuristics can nevertheless often converge to the correct solution. The existing literature has explored candidate QUBO solutions for CA of the latter kind. These representations require $O(N^2)$ \cite{deshpande2025} or $O(KN)$ \cite{mazzei2025} binary variables, where $K$ is the maximum allowed cycle length.

After preparing a problem in a valid QUBO construction, an annealing algorithm is used to identify the optimal solution. In quantum annealing, the QUBO is mapped onto an equivalent Ising Hamiltonian acting on a collection of coupled spin-1/2 objects, where the couplings and local fields encode the quadratic cost function and the ground state encodes the optimal solution. The system is initialized in the easily prepared ground state of a strong transverse field, which is then slowly ramped down while the problem Hamiltonian is ramped up; if the evolution is sufficiently slow, the system remains near its instantaneous ground state, and a final measurement of the spins yields a low-energy, ideally optimal, solution. In this work, we evaluate the QUBO formulation using classical simulated annealing and use the resulting resource estimates to identify problem sizes suitable for future hardware quantum-annealing tests.

This work presents a novel QUBO formulation for anchored currency-arbitrage cycle search that uses fewer logical variables than prior QUBO encodings, with competitive per-variable connectivity, while more closely matching practical trading constraints. Using the resulting resource estimates, we identify problem sizes that may be suitable for future hardware quantum-annealing tests. The paper first explains the goals of a CA solution. Next, the paper introduces the QUBO formulation and its scaling properties. Then, the paper presents results using classical simulated annealing. Finally, the paper ends with a discussion of findings, future hardware annealing directions, and conclusions.

\section{Understanding the Currency Arbitrage Problem}
\begin{figure}
    \centering
    \includegraphics[width=0.6\linewidth]{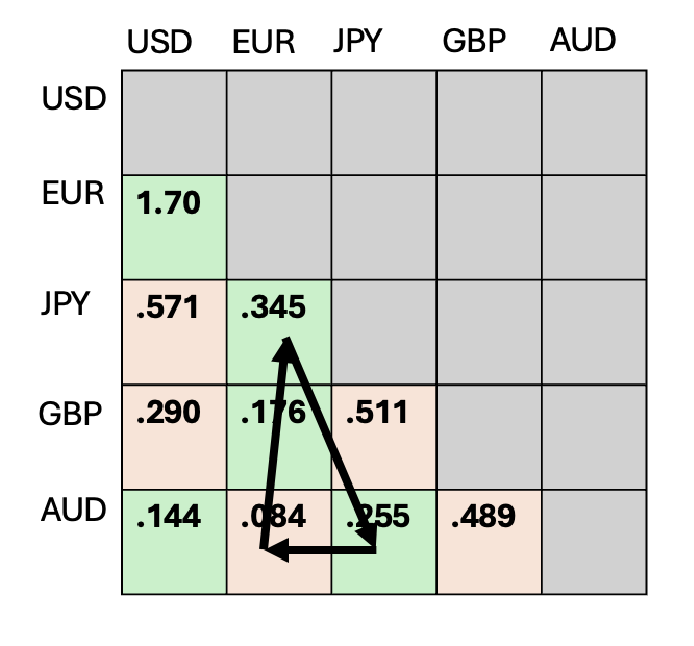}
    \caption{Extreme dummy exchange rate matrix with profit inserted. Green indicates an exchange rate was perturbed upward and orange indicates an exchange rate was perturbed downward.}
    \label{fig:exchange}
\end{figure}
The CA problem statement can be easily explained using the visual of some dummy exchange rate matrix shown in Fig.~\ref{fig:exchange}. In a perfectly balanced scenario, all currencies would trade in a way that if you traded arbitrarily many times between currencies and returned back to the original currency, you would have exactly the same amount you started with, minus trading fees. However, in the scenario shown in Fig.~\ref{fig:exchange}, minor imbalances were added by shifting some rates down in value, shown with orange squares, and some rates up in value, shown with green squares.
In this scenario, the maximal profit which can be achieved would involve trading $1\,\mathrm{AUD}\rightarrow 11.9\,\mathrm{EUR} \rightarrow 4.11\,\mathrm{JPY} \rightarrow 1.05\,\mathrm{AUD}$, leading to a $0.05\,\mathrm{AUD}$ profit per cycle if you ignore trading fees.

One existing QUBO solution for CA prioritizes the task of identifying a maximally profitable cycle regardless of cycle length \cite{deshpande2025}. Another existing solution for QUBO CA prioritizes a maximally profitable cycle with a certain upper limit of cycle length with a general reward for shorter cycles \cite{mazzei2025}. Other similar solutions have also been explored in the literature \cite{roy2025, rosenberg2016}. In real world applications of CA, there are generally additional constraints beyond just finding the maximally profitable trading cycle. One additional constraint that we aim to address in this paper is the introduction of a term that corresponds to specific fixed rate trading fees for each additional transaction. A second general constraint we introduce is that of starting from a specific currency. This is primarily due to two trading properties. First, both profit and time are lost in CA if a trade needs to be executed to enter a specific cycle. Second, it is generally preferable to begin and end trading cycles on a specific subset of reserve currencies that are more stable and profitable when held long-term during wait times between trading periods. The latter constraint has the added property of significantly reducing the number of qubits required for the solution, which, as we demonstrate later, is key to reducing logical-variable count and embedding overhead for future quantum-annealing implementations.

\section{A QUBO Solution for Currency Arbitrage}\label{sec:qubo}
The structure of the QUBO solution Hamiltonian that we introduce can be subdivided into two blocks of terms. The first block contains the constraint terms, which enforce proper construction of a single terminal cycle without reusing a currency and with only one currency per cycle step. This block we will term $H_{c}$. The second block contains all remaining terms aimed at maximizing the profitability of the cycle accounting for maximal allowed cycle length, proper closing of the cycle when profit cannot be increased, and fixed rate trading fees. This block we will term $H_{p}$.

First, let us begin with some preliminary definitions of the variables which will be used in the following equations. $s$ denotes the fixed start (anchor) currency. $C$ is the number of currencies considered and $\mathcal{C}^*$ is the set of the $C-1$ non-start currencies. $x_{i,k}\in\{0,1\}$ is a binary decision variable that indicates the insertion of a currency at a particular cycle step, with the first index corresponding to the currency and the second index corresponding to the cycle step. $S_k\equiv \sum_{i\in \mathcal{C}^*}x_{i,k}$ counts the currencies selected at step $k$. $T$ is the horizon length, equal to $K-1$ where $K$ is the maximal cycle length. $L_{ij}$ is the matrix of log-exchange rates between currencies, $L_{ij}=\ln R_{ij}$, where $R_{ij}$ is the number of units of currency $j$ received per unit of currency $i$.

The first constraint term prevents selecting more than one currency for a given cycle step and is written as follows:

\begin{equation}\label{eq:hcol}
    H_{col}=2A_{col}\sum_{k=1}^T\sum_{\substack{i<j \\ i,j\in \mathcal{C}^*}}x_{i,k}x_{j,k},
\end{equation}
where $A_{col}$ is the weight for this constraint term (the factor of two arises from writing the penalty as $A_{col}\sum_k S_k(S_k-1)$). The next constraint term prevents reusing a currency once it is already used and is written as:

\begin{equation}\label{eq:hrow}
    H_{row}=A_{row}\sum_{i\in \mathcal{C}^*} \sum_{1 \leq k< k'\leq T}x_{i,k}x_{i,k'},
\end{equation}
where $A_{row}$ is again a term weight. An additional contiguity constraint is needed in this representation to enforce that, once a cycle is closed, a new cycle is not restarted at a later step. The contiguity term is written as:

\begin{equation}\label{eq:hcontig}
    H_{contig}=A_{contig}\sum_{k=1}^{T-1}(S_{k+1}-S_{k+1}S_k),
\end{equation}
with $A_{contig}$ as the term weight. The last constraint term ensures that cycle generation starts at the first cycle step, which breaks degeneracies and reduces local minima, and is written as:

\begin{equation}\label{eq:hstart}
    H_{start}=A_{start}(1-S_1),
\end{equation}
where $A_{start}$ is the term weight. Altogether the constraint terms sum to give the total constraint term:
\begin{equation}\label{eq:hc}
    H_{c} = H_{row} + H_{col} + H_{contig} + H_{start}.
\end{equation}

The profit terms require fewer unique term weights, with a single weight $\alpha$ shared across all terms except the length penalty term, which uses the constant $\gamma$. The first profit term directly encodes the profit made on the first step of the cycle from the start currency $s$. The first step term is written as:
\begin{equation}\label{eq:hstep1}
    H_{step1} = \sum_{j\in \mathcal{C}^*}(-\alpha L_{sj})x_{j,1}.
\end{equation}
The next profit term allows a cycle, once begun, to be closed by returning to the start currency after any step. Note that closing immediately after the first step corresponds to a single back-and-forth trade, $s \rightarrow c_1 \rightarrow s$; when the exchange rates are symmetric ($L_{ij}=-L_{ji}$), such a round trip has exactly zero log-return, so permitting it is harmless. The exception to this would be cases where the trading values are dynamically changing, which we do not consider in this work. The cycle closing term is written as:
\begin{equation}\label{eq:hclose}
    H_{close} = \sum_{k=1}^{T-1}\sum_{i\in \mathcal{C}^*}(-\alpha L_{is})x_{i,k}.
\end{equation}
The other term needed when considering closing or continuing a cycle is the profit made by continuing to the next step which is encoded in the following term:
\begin{equation}\label{eq:hcontinue}
    H_{continue} = \sum_{k=1}^{T-1}\sum_{i\in \mathcal{C}^*}\sum_{j\in \mathcal{C}^*}(-\alpha(L_{ij}-L_{is}))x_{i,k}x_{j,k+1}.
\end{equation}
Terms with $i=j$ never contribute on feasible assignments, since Eq.~\eqref{eq:hrow} forbids selecting the same currency at consecutive steps.
To account for trading latencies placing limits on the maximum number of trades which can be executed, it is critical to be able to control a maximal cycle length. The horizon point, $T$, can be treated as a final step after which the loop is closed back to the start currency. The term corresponding to this truncation is written as:
\begin{equation}\label{eq:htruncate}
    H_{truncate} = \sum_{i\in \mathcal{C}^*}(-\alpha L_{is})x_{i,T}.
\end{equation}
The final profit maximization term is a length penalty term which represents a flat rate trading fee per additional step added to the cycle written as:
\begin{equation}\label{eq:hlen}
    H_{len} = \gamma \sum_{k=1}^T\sum_{i\in \mathcal{C}^*}x_{i,k}.
\end{equation}
These five terms sum to give the total profit term written as:
\begin{equation}\label{eq:hp}
    H_{p} = H_{step1} + H_{close} + H_{continue} + H_{truncate} + H_{len}.
\end{equation}

The sum of all constraints and all profits gives the final Hamiltonian for this problem:
\begin{equation}\label{eq:htotal}
    H = H_{c} + H_{p}.
\end{equation}

\noindent\textbf{Correctness of the encoding.}
For sufficiently large constraint weights \(A_{\mathrm{row}}\), \(A_{\mathrm{col}}\), \(A_{\mathrm{contig}}\), and \(A_{\mathrm{start}}\), minimizing the full Hamiltonian restricts the lowest-energy solutions to feasible assignments of the constraint Hamiltonian in Eq.~\eqref{eq:hc}. For any such feasible assignment, the selected binary variables correspond to a contiguous, non-repeating cycle anchored at the fixed starting currency \(s\),

\[ s \rightarrow c_1 \rightarrow c_2 \rightarrow \cdots \rightarrow c_m \rightarrow s, \]

where \(1\leq m\leq T\), \(x_{c_k,k}=1\) for \(k=1,\ldots,m\), and all later cycle positions are empty. The first-step term in Eq.~\eqref{eq:hstep1} contributes the starting exchange contribution \(-\alpha L_{s c_1}\). For each occupied nonterminal position \(k<m\), the closing term in Eq.~\eqref{eq:hclose} first assigns a provisional return-to-start contribution \(-\alpha L_{c_k s}\). If the cycle continues to \(c_{k+1}\), the continuation term in Eq.~\eqref{eq:hcontinue} contributes

\[ -\alpha\left(L_{c_k c_{k+1}}-L_{c_k s}\right), \]

which cancels the provisional closing contribution and replaces it with the continuation-edge contribution \(-\alpha L_{c_k c_{k+1}}\). At the terminal occupied position, no following currency is selected, so the provisional closing contribution remains when \(m<T\). If the cycle reaches the horizon, \(m=T\), the final closing contribution is instead supplied by the truncation term in Eq.~\eqref{eq:htruncate}. Thus, for any feasible cycle, the profit Hamiltonian evaluates to

\begin{equation}\label{eq:hpfeasible}
H_p = -\alpha\left(L_{s c_1} + \sum_{k=1}^{m-1} L_{c_k c_{k+1}} + L_{c_m s} \right) + \gamma m.
\end{equation}

Because the sum of log-exchange rates is the log-return of the full arbitrage cycle, minimizing the above expression over feasible assignments is equivalent to maximizing the cycle log-return while applying a fixed length penalty of \(\gamma/\alpha\) per selected non-start currency. This is equivalent to a fixed per-transaction penalty up to a constant offset associated with the final return-to-start exchange, which does not affect the minimizing cycle among the nonempty cycles enforced by $H_{start}$. This penalty has a direct financial reading. If each executed trade delivers a fraction $1-f$ of its nominal proceeds, then choosing $\gamma/\alpha=-\ln(1-f)\approx f$ makes $H_{len}$ charge a uniform proportional fee on $m$ of the $m+1$ trades. The uncharged final return-to-anchor trade is the constant offset described above. Pair-specific fee schedules $f_{ij}$ can instead be absorbed directly into the log-rates as $L_{ij}\rightarrow L_{ij}+\ln(1-f_{ij})$. This charges all $m+1$ trades and reduces $H_{len}$ to a pure length regularizer. In all experiments with $\gamma>0$ in this work, the exact baseline optimizes the same fee-adjusted objective, $\sum L-(\gamma/\alpha)\,m$. Therefore, minimizing the full Hamiltonian gives the desired anchored arbitrage cycle, provided the constraint weights are chosen large enough that infeasible assignments are never energetically favored over feasible cycles. Explicit sufficient conditions are derived next.

\subsection{Sufficient penalty weights}\label{sec:weights}
Existing QUBO encodings for CA generally leave the constraint weights unspecified or tuned empirically. In this section, we provide explicit sufficient conditions for the weights. We define $L_\infty=\max_{a\neq b}|L_{ab}|$ over the log-rate matrix actually supplied to the QUBO. We also assume $T\leq N-1$, which holds without loss of generality because a simple anchored cycle visits at most $N-1$ non-start currencies.

\medskip
\noindent\textbf{Theorem (sufficient penalty weights).} \emph{If}
\begin{align}
A_{start}&> \gamma+2\alpha L_\infty,\label{eq:bound_s}\\
A_{contig}&> (T-1)\,\max\!\left(0,\ \alpha L_\infty-\gamma\right),\label{eq:bound_c}\\
A_{row}&> 4\alpha L_\infty,\label{eq:bound_r}\\
A_{col}&> 2A_{contig}+5\alpha L_\infty+\tfrac{1}{2}A_{start},\label{eq:bound_q}
\end{align}
\emph{then every ground state of $H$ in Eq.~\eqref{eq:htotal} encodes a nonempty, anchored, contiguous, simple cycle, and hence, by Eq.~\eqref{eq:hpfeasible}, a maximizer of the fee-adjusted log-return $\sum L - (\gamma/\alpha)m$.}

\medskip
The proof works by showing that each class of constraint violation admits a local repair move which strictly lowers the energy. The moves are ordered so that each step relies only on the violation classes already excluded. As a consequence, no infeasible configuration is a local minimum under these moves, which is a useful property for annealing.

First, for Eq.~\eqref{eq:bound_q}, suppose some cycle step is multi-occupied. Deleting one variable from a maximally occupied step with occupancy $M\geq2$ changes the energy by at most $-2A_{col}(M-1)+2A_{contig}M+A_{start}+2\alpha L_\infty+4\alpha L_\infty M$, and the case $M=2$ gives the bound. The weights $A_{contig}$ and $A_{start}$ appear inside this bound because both $H_{contig}=A_{contig}\sum_k S_{k+1}(1-S_k)$ and $H_{start}=A_{start}(1-S_1)$ become negative on multi-occupied configurations. These terms act as rewards for multi-occupancy which the deletion forfeits.

Next, for Eq.~\eqref{eq:bound_r}, suppose all steps are singly occupied and some currency occurs at steps $k_1<k_2$. We swap the occurrence at step $k_2$ for an unused currency, which exists whenever $T\leq N-1$. The telescoping between Eq.~\eqref{eq:hclose} and Eq.~\eqref{eq:hcontinue} reduces the energy contribution of the currency $b$ at step $k_2$ to $-\alpha(L_{\mathrm{prev},b}+L_{b,\mathrm{next}})$ plus terms independent of $b$, so the swap changes at most two log-rate entries. This gives the bound of $4\alpha L_\infty$. The same argument covers the $i=j$ couplings of Eq.~\eqref{eq:hcontinue}. These couplings vanish on feasible assignments, but on infeasible assignments they shift the effective consecutive-reuse penalty to $A_{row}+\alpha L_{i s}$, which Eq.~\eqref{eq:bound_r} absorbs.

Then, for Eq.~\eqref{eq:bound_c}, consider a detached segment, meaning a maximal occupied run of $\ell$ steps that does not contain the first step. Such a segment pays exactly one $A_{contig}$ penalty. Its profit terms telescope to the log-return of a phantom path of $\ell$ edges that ends at the anchor without ever leaving it. Deleting the whole segment changes the energy by $-A_{contig}+\alpha(\text{path return})-\gamma\ell\leq -A_{contig}+\ell(\alpha L_\infty-\gamma)$ with $\ell\leq T-1$, which gives the bound. This bound is tight in the worst case, since an adversarial rate matrix can hide a path with a return of order $\ell L_\infty$. The data-dependent quantity that actually matters is the maximum phantom-path return, which is small for near-balanced markets.

Finally, for Eq.~\eqref{eq:bound_s}, the all-empty configuration costs exactly $A_{start}$, while the best feasible cycle costs at most $\gamma-\alpha\max_j(L_{sj}+L_{js})\leq\gamma+2\alpha L_\infty$.

All four conditions are homogeneous of degree one in $(\alpha,\gamma)$. The parameter $\alpha$ is therefore a free positive scale, only the ratios $A/(\alpha L_\infty)$ and $\gamma/\alpha$ matter, and $\alpha$ can be chosen to place the final coefficient range within hardware limits. We verified the theorem empirically by exhaustively enumerating the full $2^{(N-1)T}$-state spectrum on small instances with general asymmetric rate matrices, with and without trading fees and gauge fixing. Under the derived weights, the ground state coincided with the independently computed Held--Karp optimum in every case. The simulated-annealing sweeps of Sec.~\ref{sec:sim} use the heuristic weights $A_{col}=2\alpha L_\infty+\gamma$, $A_{row}=2\alpha L_\infty$, $A_{contig}=\alpha L_\infty$, and $A_{start}=A_{col}$. These fall below the worst-case bounds above, and for near-balanced data they sit at the boundary set by the phantom-path return. The encoding comparison of Sec.~\ref{sec:encoding_comparison} uses the derived weights.

\subsection{Gauge fixing and coefficient dynamic range}\label{sec:gauge}
The bounds above are all linear in $L_\infty$. For realistic exchange rates, $L_\infty$ is set by the overall scale of the log-rates, which is of order unity, while the profits of interest live at the arbitrage scale of $10^{-4}$. This mismatch is the principal obstacle to resolving basis-point profits on analog annealing hardware with finite coupling precision. It can be removed exactly. We replace
\begin{equation}\label{eq:gauge}
L_{ij}\ \rightarrow\ \tilde{L}_{ij}=L_{ij}-(\varphi_j-\varphi_i),\qquad \varphi_i=L_{si},\ \varphi_s=0,
\end{equation}
which is a Johnson-style \cite{johnson77} reweighting by node potentials. Every closed cycle's log-return is invariant under Eq.~\eqref{eq:gauge} because the potentials telescope, so the optimal cycle is unchanged. After the transformation, $\tilde{L}_{sj}=0$ identically, $\tilde{L}_{is}=L_{is}+L_{si}$ is the round-trip spread through the anchor, and $\tilde{L}_{ij}=L_{ij}-L_{sj}+L_{si}$ is the triangular-arbitrage deviation through the anchor. All of these live at the arbitrage scale. Because every bound in Eqs.~\eqref{eq:bound_s}--\eqref{eq:bound_q} is linear in $L_\infty$, the derived weights shrink by the same factor. On the ten-currency, $K=6$ benchmark instance used below, gauge fixing reduces the largest QUBO coefficient from $25.3$ to $1.9\times10^{-2}$. It also reduces the ratio of the largest coefficient to the optimal fee-adjusted log-return from roughly $3.6\times10^{4}$ to $26$, bringing the required analog precision within the reach of current annealers. Gauge fixing acts only on the input rate matrix, so it applies as preprocessing to any cycle-based CA encoding. In Sec.~\ref{sec:encoding_comparison}, we show that it improves every prior formulation as well. Because the potentials are taken from the anchor row, the transformation and the derived weights are recomputed for each anchor.

\begin{figure}
    \centering
    \begin{tabular}{@{} c | c @{}}
        \subcaptionbox{Clique connections of CA solution.\label{fig:clique}}
        {\includegraphics[width=.5\linewidth]{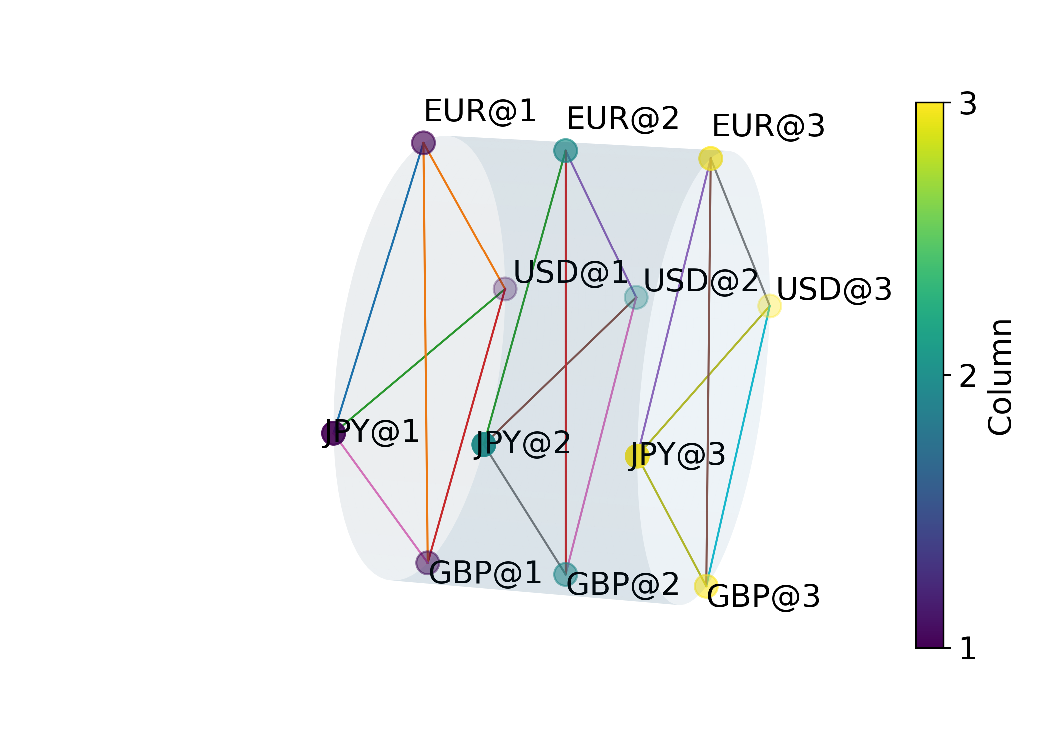}}
        &
        \subcaptionbox{Adjacent connections of CA solution.\label{fig:adjacent}}
        {\includegraphics[width=.5\linewidth]{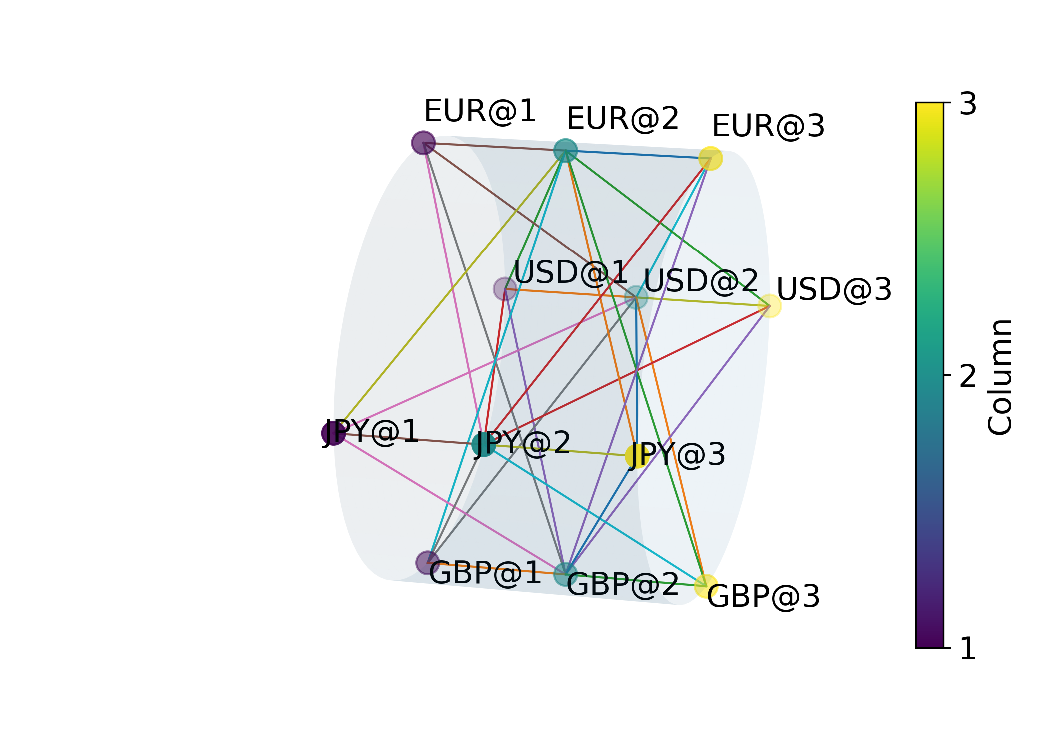}}        
    \end{tabular}
    \caption{Decomposition of the connectivity graph of the QUBO for the example in Fig.~\ref{fig:exchange}: column-clique couplings (a) and adjacent-step couplings (b).} 
\end{figure}

\begin{figure}
    \centering
    \begin{tabular}{@{} c | c @{}}
        \subcaptionbox{Row uniqueness connections of CA solution.\label{fig:row}}
        {\includegraphics[width=.5\linewidth]{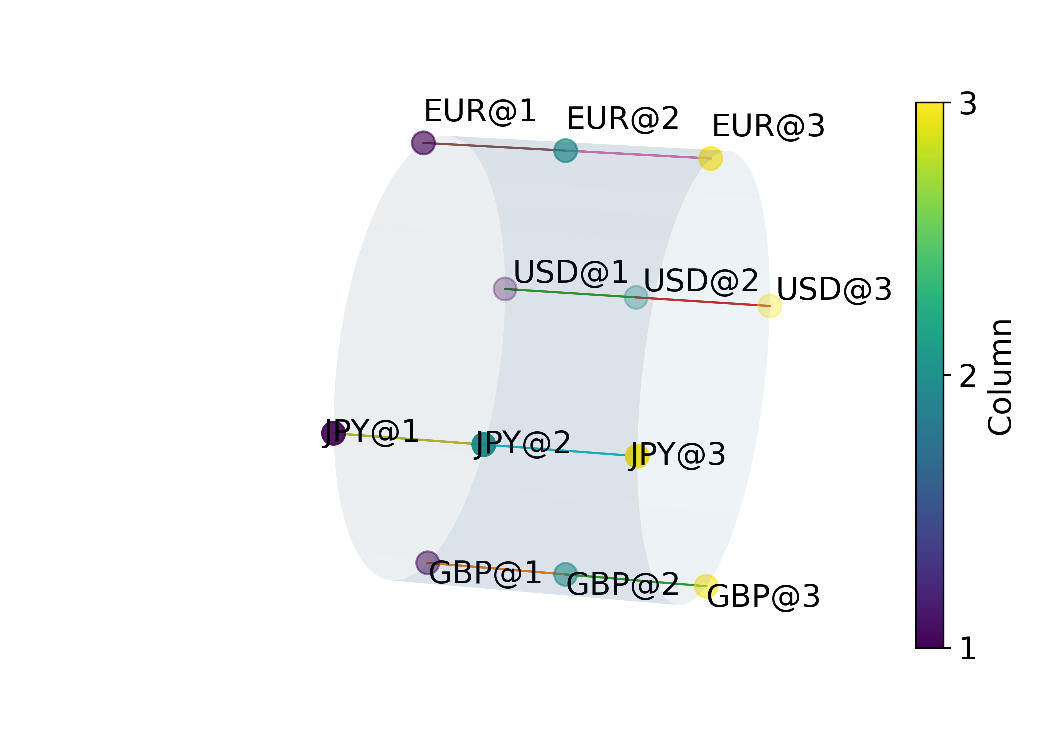}}
        &
        \subcaptionbox{Full graph connectivity of CA solution.\label{fig:full}}
        {\includegraphics[width=.5\linewidth]{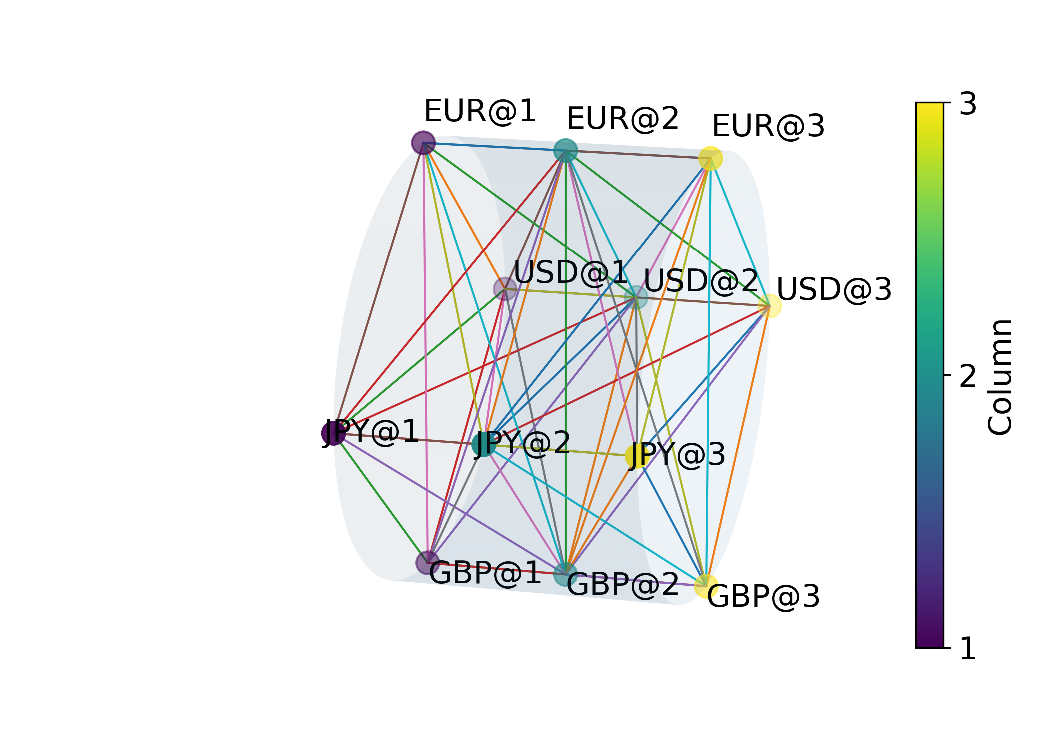}}        
    \end{tabular}
    \caption{Decomposition of the connectivity graph (continued): row-uniqueness couplings (a) and the full coupling graph (b).}
\end{figure}

The connectivity of the graph associated with the example in Fig.~\ref{fig:exchange} can be broken down into clique connections shown in Fig.~\ref{fig:clique}, row uniqueness connections shown in Fig.~\ref{fig:row}, and adjacency connections shown in Fig.~\ref{fig:adjacent}, with the total connected graph shown in Fig.~\ref{fig:full}.

A comparison of connectivity with existing solutions in the literature is provided in Table~\ref{tab:scaling}. The number of logical qubits for this novel solution is less than that of the existing solutions in the literature. All of these existing prior solutions could have $N$ replaced by $N-1$ by using a similar approach of anchoring to a fixed start position, which more realistically reflects real-world strategies. Among the solutions which scale like $O(NK)$ in number of qubits, as opposed to $O(N^2)$, this solution has a maximal single-qubit connectivity lower than or equal to that of the alternatives for maximum cycle lengths $K\leq7$; for $K\geq8$, the $K$-independent maximal connectivity of Mazzei et al.\ \cite{mazzei2025} ($3N-1$) becomes lower, although that encoding requires more logical qubits. The logical-qubit-count scaling of this novel solution is preferable to previous solutions in the literature.

\begin{table}
    \centering
    \begin{tabular}{ccc}
        \toprule
        Encoding & Logical Qubits & Maximal Connectivity \\
        \midrule
        Mazzei et al. & \multirow{2}{*}{$NK$} & \multirow{2}{*}{$3N-1$}\\
        \cite{mazzei2025} & & \\
        \midrule
        Deshpande–Das–Mueller & \multirow{2}{*}{$N^2$} &  \multirow{2}{*}{$2N-4$} \\
        \cite{deshpande2025} & & \\
        \midrule
        Roy et al. & \multirow{2}{*}{$N(K+1)$} & \multirow{2}{*}{$3N+K-3$}\\
        \cite{roy2025} & & \\
        \midrule
        1QBit (node-based) & \multirow{2}{*}{$NK$} & \multirow{2}{*}{$3N+K-4$} \\
        \cite{rosenberg2016} & & \\
        \midrule
        1QBit (edge-based) & \multirow{2}{*}{$N^2$} & \multirow{2}{*}{$4N - 7$} \\
        \cite{rosenberg2016} & & \\
        \midrule
        \textbf{Ours} & $(N-1)(K-1)$ & $3N+K-8$ \\
        \bottomrule
    \end{tabular}
    \caption{Logical qubit count and maximal connectivity of the single most-connected qubit for existing CA QUBO solutions in the literature.}
    \label{tab:scaling}
\end{table}

\begin{figure}
    \centering
    \begin{tabular}{@{} c | c @{}}
        \subcaptionbox{Number of fully-connected qubits vs number of currencies for a maximum cycle length of 6.\label{fig:fc_currencies}}
        {\includegraphics[width=.5\linewidth]{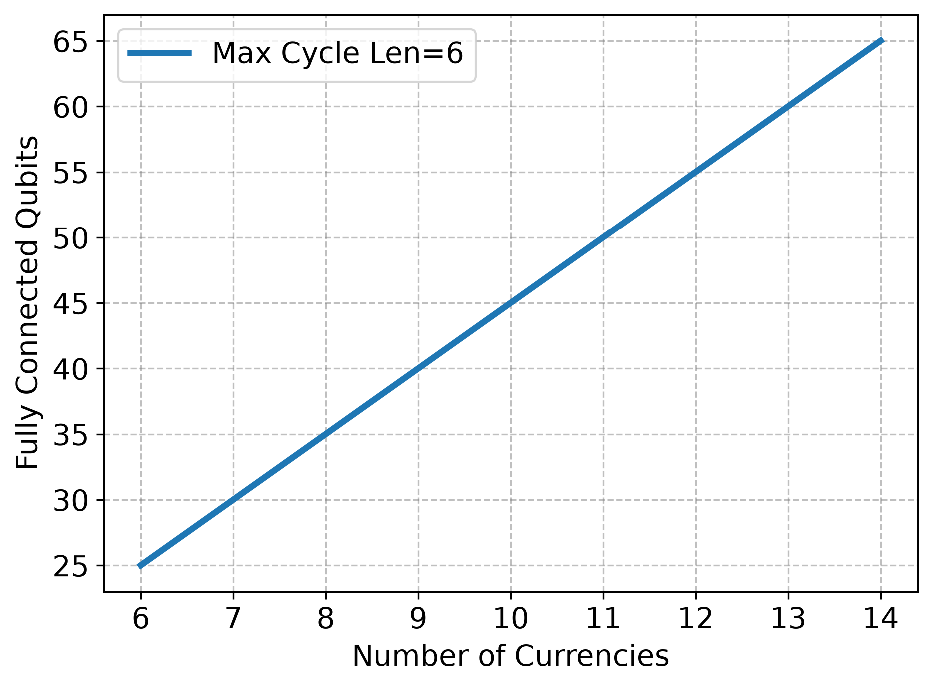}}
        &
        \subcaptionbox{Number of qubits after embedding onto the Pegasus-16 graph vs number of currencies for a maximum cycle length of 6.\label{fig:pg16_currencies}}
        {\includegraphics[width=.5\linewidth]{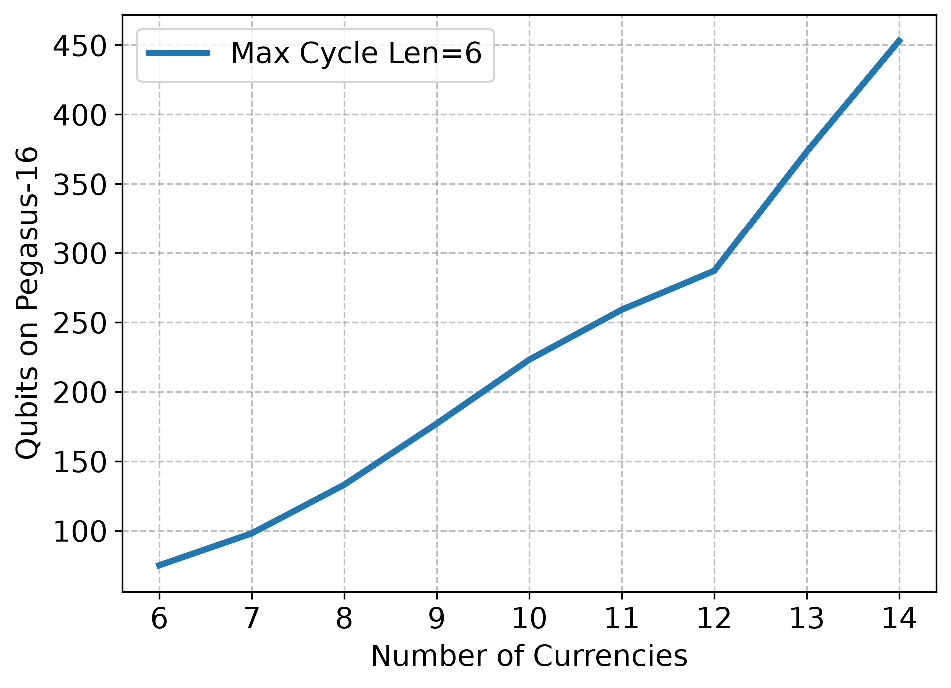}}        
    \end{tabular}
    \caption{Logical qubit count (a) and physical qubit count after minor-embedding onto the Pegasus-16 graph (b) as a function of the number of currencies, for a maximum cycle length of six.}
\end{figure}

\begin{figure}
    \centering
    \begin{tabular}{@{} c | c @{}}
        \subcaptionbox{Number of fully-connected qubits vs maximum cycle length for 14 currencies.\label{fig:fc_length}}
        {\includegraphics[width=.45\linewidth]{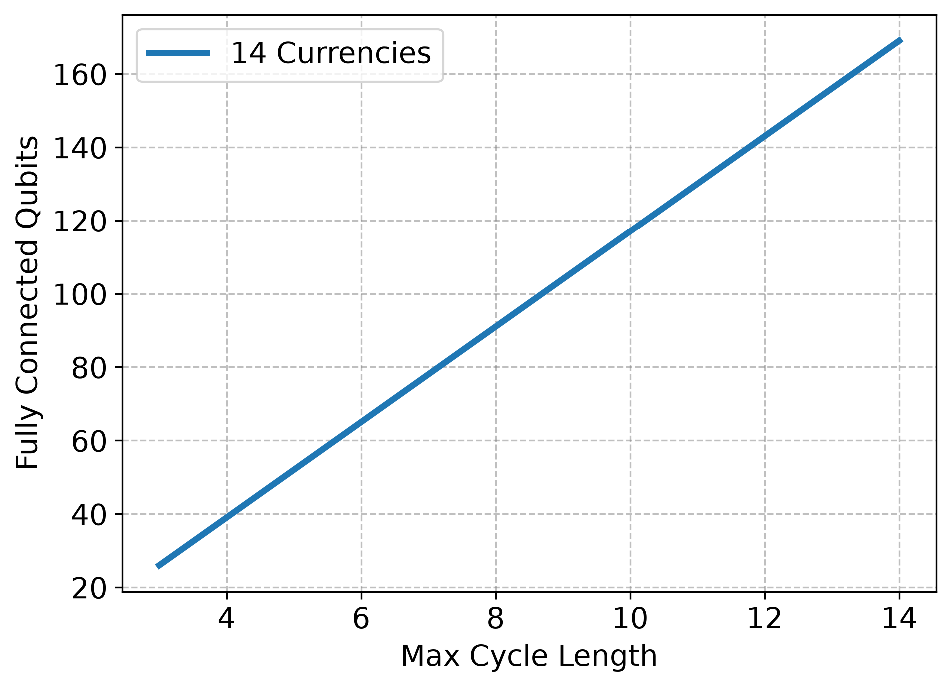}}
        &
        \subcaptionbox{Number of qubits after embedding onto the Pegasus-16 graph vs maximum cycle length for 14 currencies.\label{fig:pg16_length}}
        {\includegraphics[width=.45\linewidth]{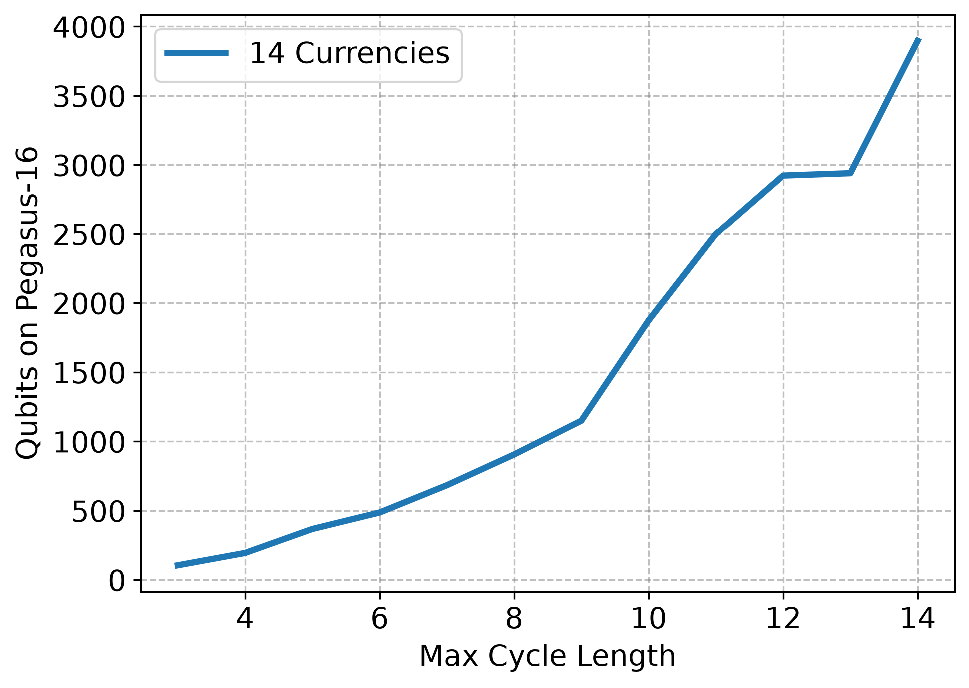}}        
    \end{tabular}
    \caption{Logical qubit count (a) and physical qubit count after minor-embedding onto the Pegasus-16 graph (b) as a function of maximum cycle length, for 14 currencies.}
\end{figure}

An additional note about scaling behavior of these algorithms for use on hardware quantum annealers is that the maximal connectivity described in Table~\ref{tab:scaling} generally exceeds that of modern annealing hardware. For example, the D-Wave Advantage quantum annealer has native connectivity of degree 15 on the Pegasus-16 architecture. To evaluate how these algorithms are likely to scale to such systems, it is necessary to compare the number of logical variables with the number of physical qubits required after minor-embedding onto the Pegasus-16 graph. In Fig.~\ref{fig:fc_currencies} and Fig.~\ref{fig:pg16_currencies}, we compare the number of logical variables with the number of physical qubits required after embedding onto the Pegasus-16 graph. We show similar results in Fig.~\ref{fig:fc_length} and Fig.~\ref{fig:pg16_length} for qubit scaling as a function of maximum cycle length. The embedding analysis indicates that systems up to 17 currencies with a maximum cycle length of 14 can fit on the D-Wave Advantage architecture, requiring 4,343 physical qubits with a maximum chain length of 40 and an average chain length of approximately 20.88. A naive exact-enumeration baseline would require checking over 59 trillion permutations in this case, although we emphasize that substantially faster exact classical methods exist at this scale, as discussed in Sec.~\ref{sec:sim}.

\section{Results of Simulated Annealing for Currency Arbitrage}\label{sec:sim}
In order to evaluate the efficacy of the algorithm, we use D-Wave's classical \texttt{SimulatedAnnealingSampler}, which implements, by default, single-spin Metropolis updates \cite{metropolis53} swept over the variables under a decreasing temperature schedule. Energy-decreasing spin flips are always accepted, while energy-increasing flips are accepted with a Boltzmann probability, allowing escape from local minima. Convergence to the global optimum is guaranteed only for impractically slow cooling schedules, so for practical schedules the algorithm may return suboptimal solutions on rugged energy landscapes such as the one considered here.
We also break from the aforementioned real-world use case, where the algorithm would be run with a single fixed starting currency in mind, and instead iterate once for each currency as the starting currency to probe the upper limit of what the algorithm can capture.

As a benchmark comparison, we use a Held--Karp-style dynamic program over (visited subset, last currency) states, which solves the anchored, bounded-length, simple-cycle problem exactly in $O(2^{N-1}N^2)$ time and $O(2^{N-1}N)$ memory per anchor, optimizing the identical fee-adjusted objective $\sum L-(\gamma/\alpha)m$ over the identical feasible set (including single-hop round trips). At the problem sizes considered here, the dynamic program is faster by many orders of magnitude, and using a weak exact baseline would overstate any annealing advantage.

We also note that D-Wave's \texttt{SimulatedAnnealingSampler} is a compiled CPU implementation. All wall-time comparisons in this section are therefore CPU-versus-CPU on the same machine. Reported annealing times measure the sampler call only, summed over the $N$ anchor runs, with model construction timed separately.

\begin{figure}
    \centering
    \begin{tabular}{@{} c | c @{}}
        \subcaptionbox{Profit vs simulated-annealing steps for ten currencies with a maximum cycle length of six. A stable range is shown from 80 to 320 steps.\label{fig:profit_vs_steps}}
        {\includegraphics[width=.5\linewidth]{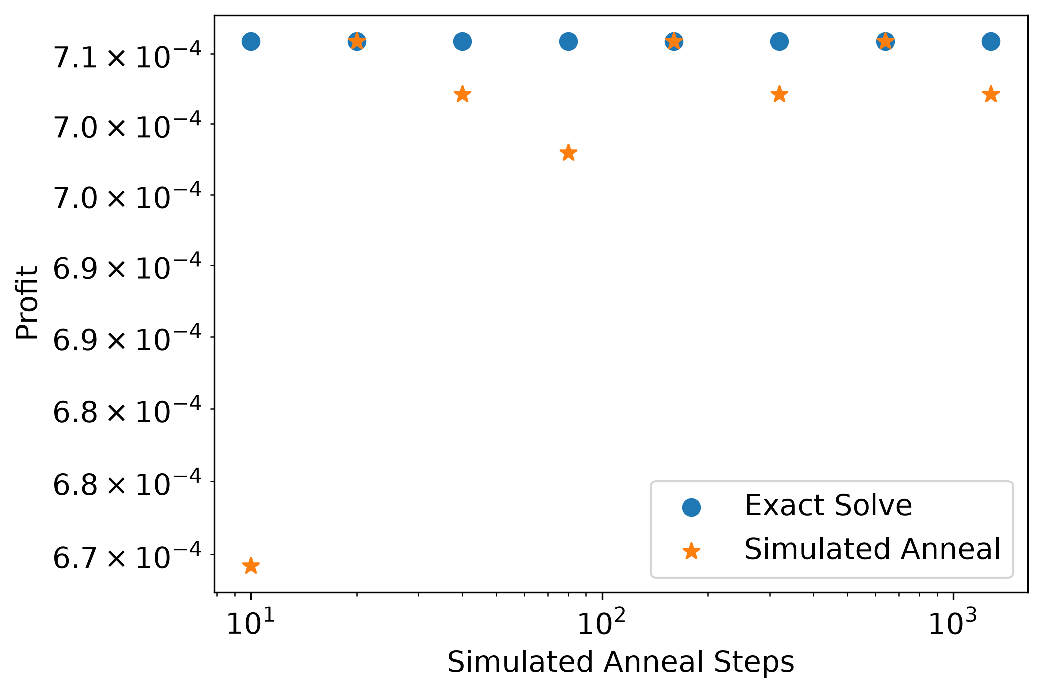}}
        &
        \subcaptionbox{Profit vs number of reads for ten currencies with a maximum cycle length of six. Results are strong at 1280 reads and above.\label{fig:profit_vs_shots}}
        {\includegraphics[width=.5\linewidth]{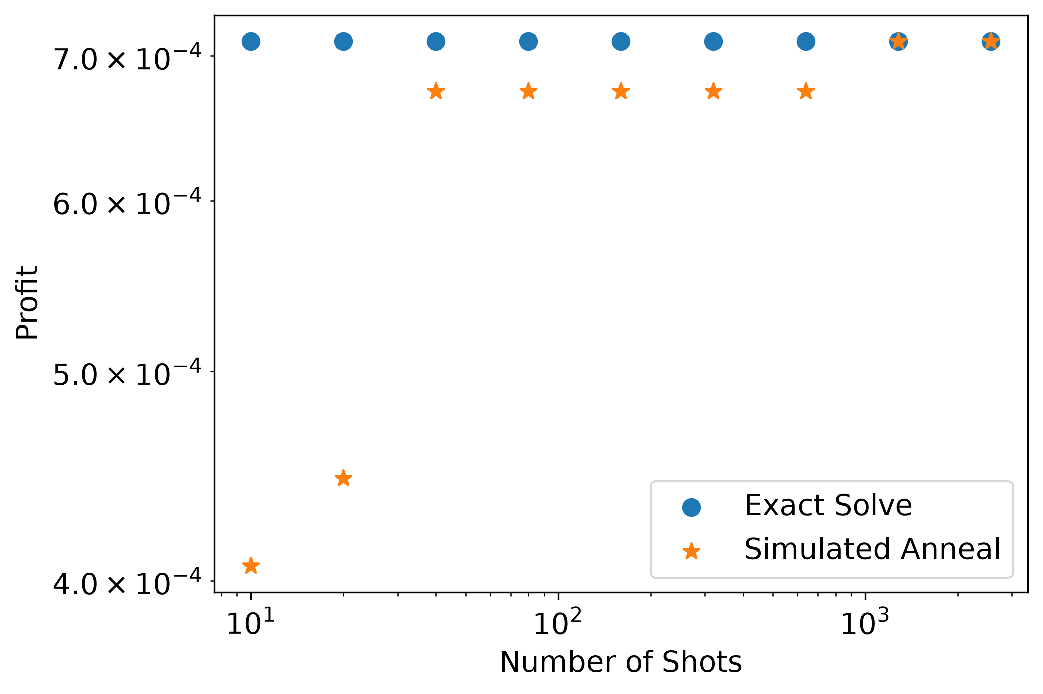}}        
    \end{tabular}
    \caption{Simulated-annealing hyperparameter sweeps for ten currencies with a maximum cycle length of six: profit vs annealing steps (a) and profit vs number of reads (b).}
\end{figure}

\begin{figure}
    \centering
    \includegraphics[width=0.5\linewidth]{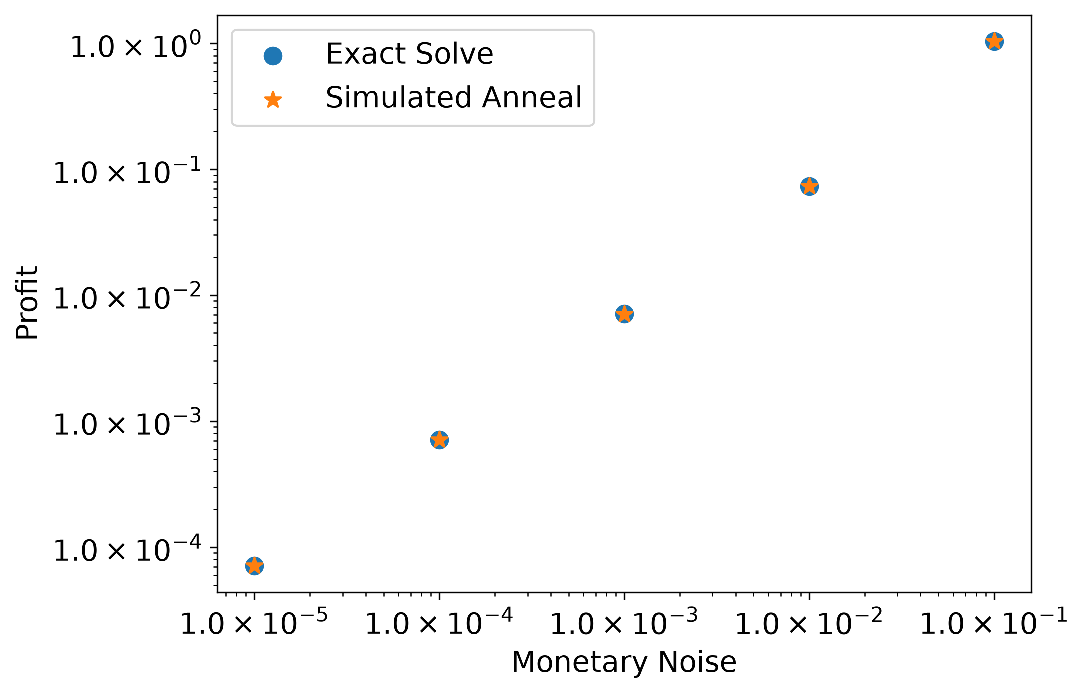}
    \caption{Profit vs monetary noise injected into the balanced currency matrix with ten currencies and a maximum cycle length of 6.}
    \label{fig:profit_noise}
\end{figure}

\begin{figure}
    \centering
    \begin{tabular}{@{} c | c @{}}
        \subcaptionbox{Profit vs number of currencies with a maximum cycle length of six and 1280 reads. The simulated annealer gives degraded results above eleven currencies.\label{fig:profit_currencies_short}}
        {\includegraphics[width=.5\linewidth]{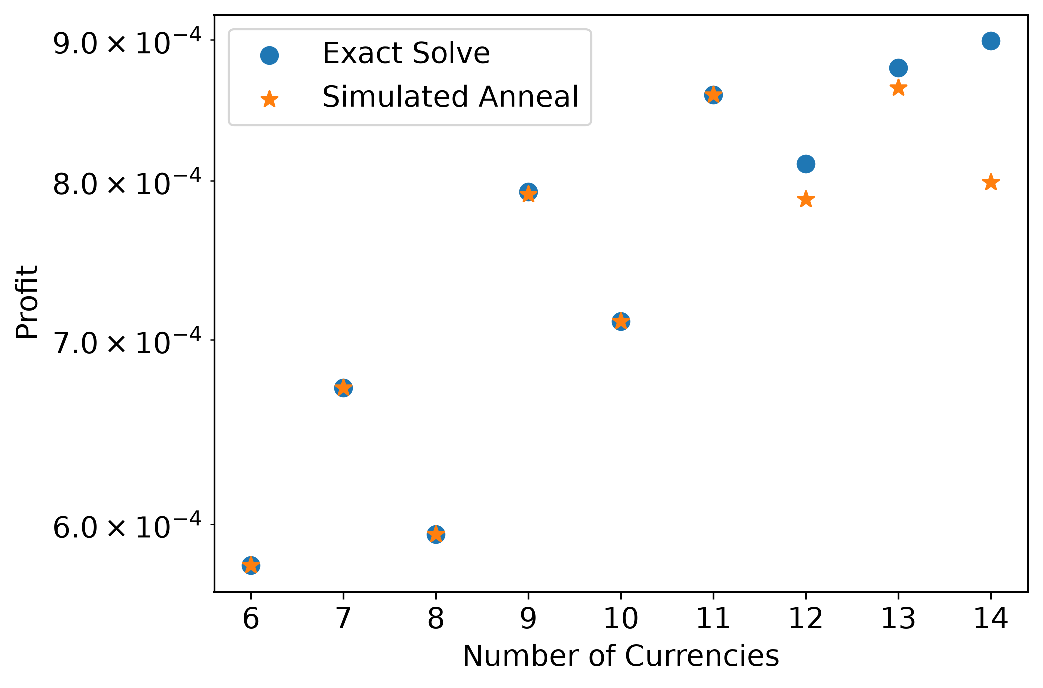}}
        &
        \subcaptionbox{Profit vs number of currencies with a maximum cycle length of six and 10240 reads. The simulated annealer recovered the exact optimum for all tested instances except $N=14$.\label{fig:profit_currencies_long}}
        {\includegraphics[width=.5\linewidth]{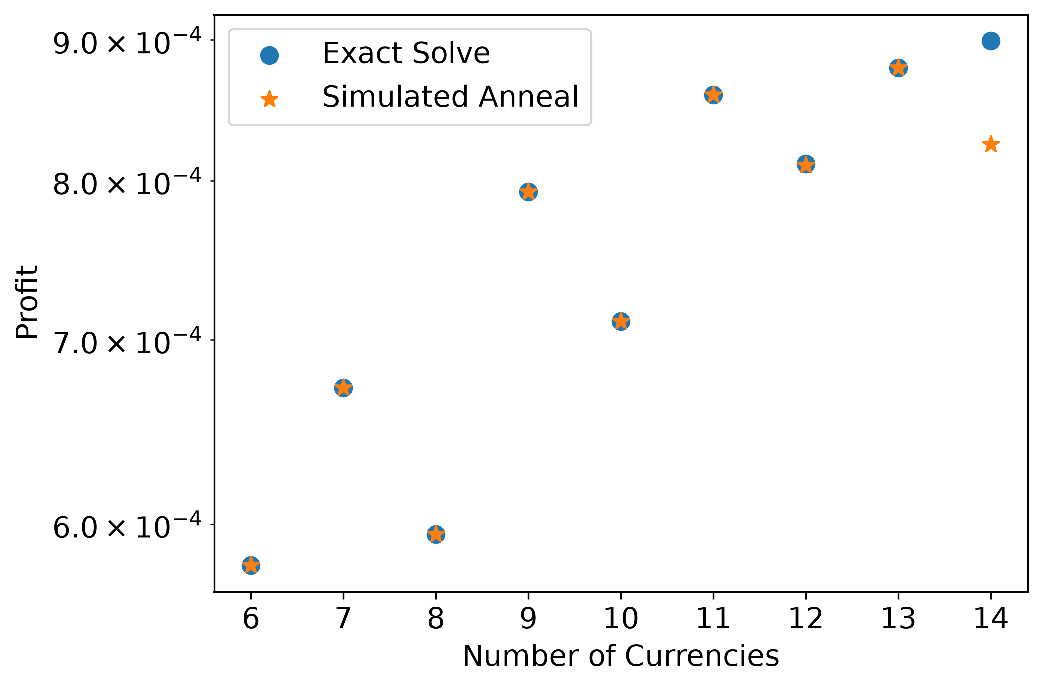}}        
    \end{tabular}
    \caption{Simulated-annealing profit compared to the exact optimum as a function of the number of currencies, with 1280 reads (a) and 10240 reads (b).}
\end{figure}

First, we conduct a simple test with ten currencies and a maximum cycle length of six. We find, when sweeping parameter values by powers of two, that the algorithm can achieve identical profit to the exact solution for between 80 and 320 steps (Fig.~\ref{fig:profit_vs_steps}) and with at least 1280 reads (Fig.~\ref{fig:profit_vs_shots}). We also show in Fig.~\ref{fig:profit_noise}, that these results still hold for minimal market imbalances though generally we will default to a monetary noise of $1\times10^{-4}$, which results in profit scales on the order of 1/10th of a penny per cycle per ``dollar''.
However, these results do not hold for very large numbers of currencies. For twelve currencies and above, additional reads are required. We show the results with the same 1280 reads in Fig.~\ref{fig:profit_currencies_short} and with 10240 reads in Fig.~\ref{fig:profit_currencies_long}.

For these variable numbers of currencies with a maximum cycle length of six, we also show the compute-time comparison between simulated annealing and the exact Held--Karp baseline, both on the same CPU. Across the tested range, the exact solver is one to two orders of magnitude faster than simulated annealing at 1280 reads and two to three orders of magnitude faster at 10240 reads.

\begin{figure}
    \centering
    \begin{tabular}{@{} c | c @{}}
        \subcaptionbox{Compute time vs number of currencies for simulated annealing and the exact Held--Karp baseline on the same CPU, with 1280 reads.\label{fig:time_currencies_short}}
        {\includegraphics[width=.5\linewidth]{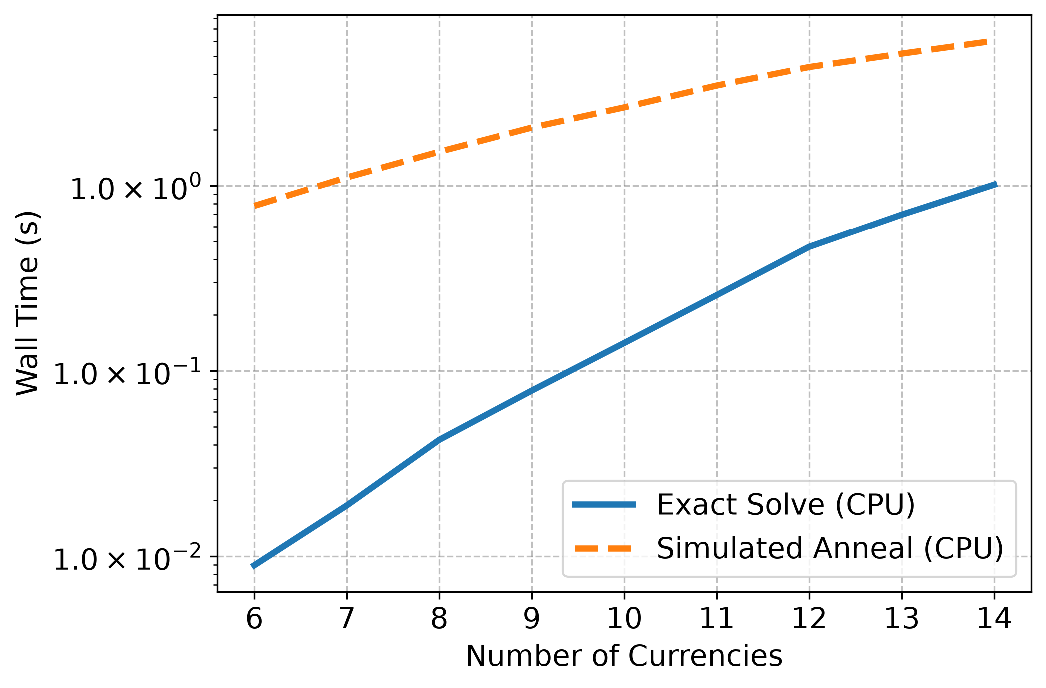}}
        &
        \subcaptionbox{Compute time vs number of currencies for simulated annealing and the exact Held--Karp baseline on the same CPU, with 10240 reads.\label{fig:time_currencies_long}}
        {\includegraphics[width=.5\linewidth]{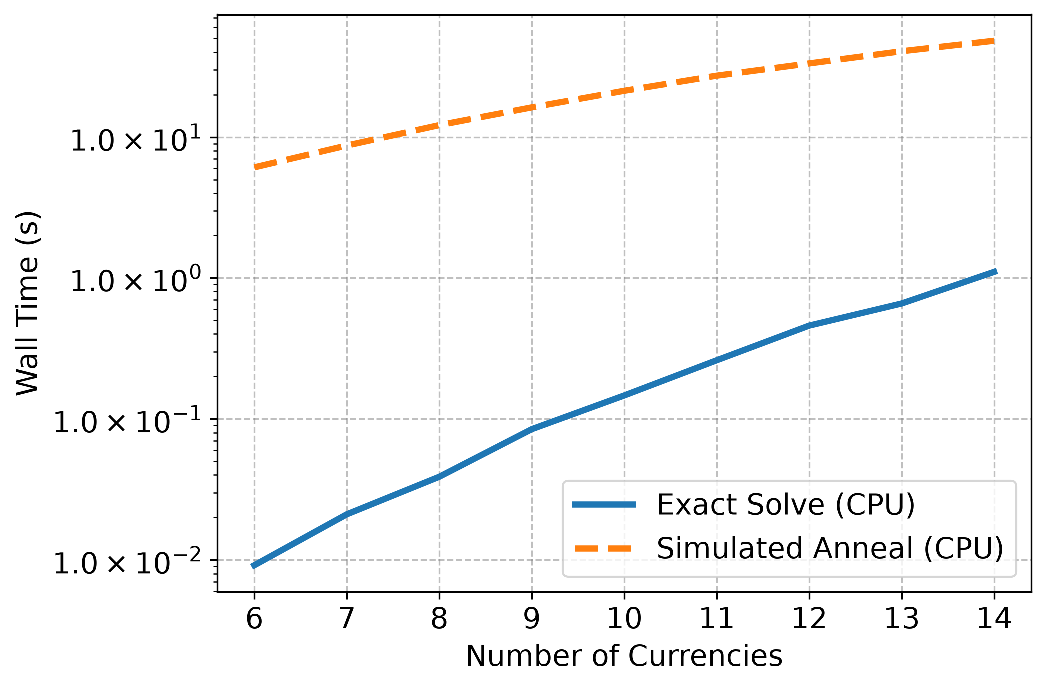}}        
    \end{tabular}
    \caption{Compute wall time for simulated annealing and the exact Held--Karp baseline, on the same CPU, as a function of the number of currencies, with 1280 reads (a) and 10240 reads (b).}
\end{figure}

\begin{figure}
    \centering
    \begin{tabular}{@{} c | c @{}}
        \subcaptionbox{Wall time vs maximum cycle length with 14 currencies and 10240 reads, on the same CPU. The exact Held--Karp baseline remains fastest throughout the tested range.\label{fig:time_length}}
        {\includegraphics[width=.5\linewidth]{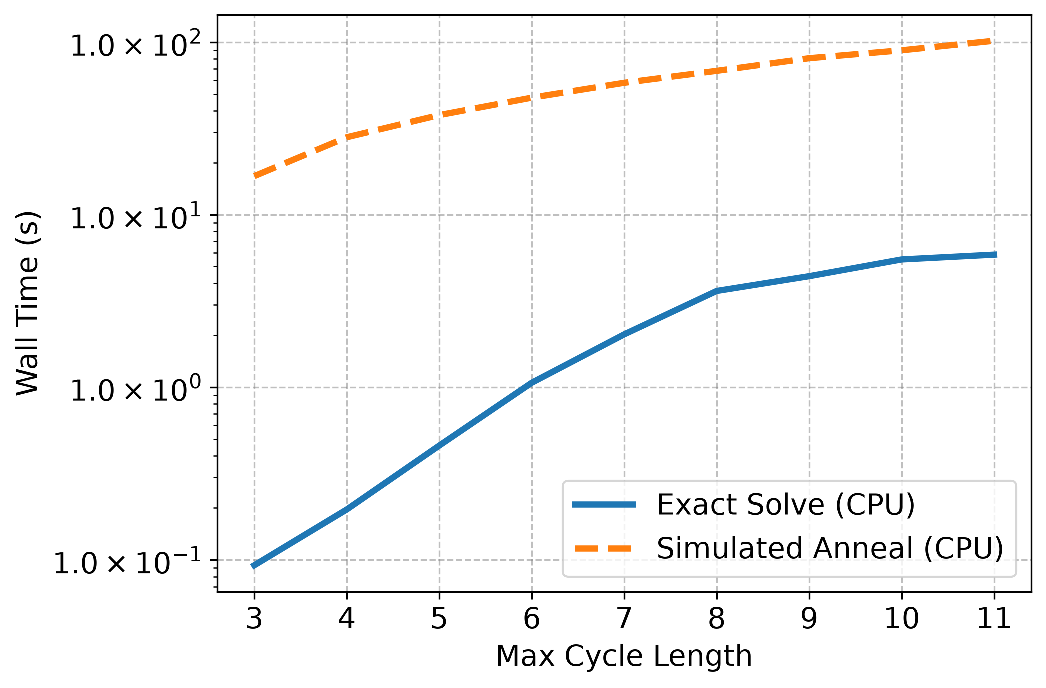}}
        &
        \subcaptionbox{Profit vs maximum cycle length with 14 currencies and 10240 reads. Profit is no longer optimal above a cycle length of five.\label{fig:profit_length}}
        {\includegraphics[width=.5\linewidth]{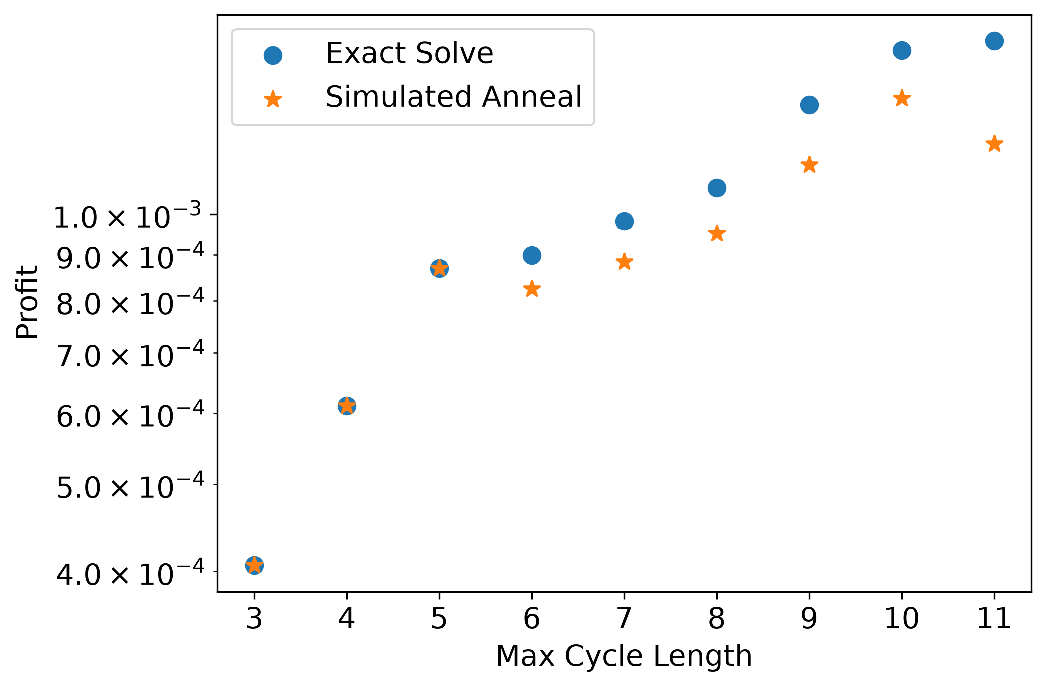}}        
    \end{tabular}
    \caption{Wall time (a) and profit (b) as a function of maximum cycle length for 14 currencies with 10240 reads.}
\end{figure}

Against this baseline, no runtime crossover is observed anywhere in the tested range. In Fig.~\ref{fig:time_length}, the exact solver remains one to two orders of magnitude faster than simulated annealing for 14 currencies at every maximum cycle length up to eleven. We show in the adjacent Fig.~\ref{fig:profit_length} that the annealed profit is no longer optimal once the maximum cycle length exceeds five. The asymptotics still favor sampling in one specific regime. For a fixed maximum cycle length $K$, the classical problem is polynomial, at $O(N^{K-1})$ by direct tuple enumeration, so no annealing speedup should be claimed there. When $K$ grows with $N$, the best exact baseline we know of is the $O(2^{N-1}N^2)$ dynamic program per anchor, which also carries an exponential memory footprint, while the anchored QUBO grows only polynomially at $(N-1)^2$ variables per anchor across $N$ anchors. On a single CPU core, the dynamic program still completes the unbounded 17-currency problem in under a minute. The classical wall therefore sits near 20 to 25 currencies. Beyond that point, exhaustive exact search becomes impractical, and sampling methods, classical or quantum, are the remaining option. Per-instance error bars and time-to-solution statistics in that regime are left to the hardware study proposed below.

\begin{figure}
    \centering
    \begin{tabular}{@{} c | c @{}}
        \subcaptionbox{Length penalty $\gamma$ vs maximum found cycle length, with the discrepancy between found and optimal cycle length as the color scale.\label{fig:gamma_length}}
        {\includegraphics[width=.45\linewidth]{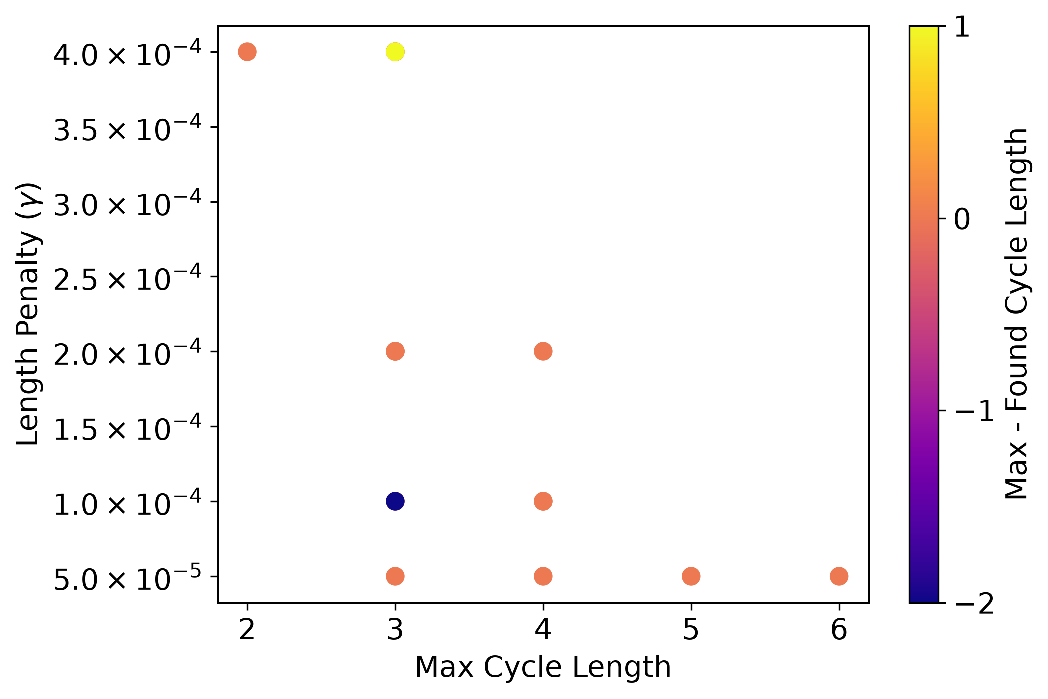}}
        &
        \subcaptionbox{Length penalty $\gamma$ vs maximum found cycle length, with the ratio of found to maximum raw profit as the color scale. Values above one and at zero are visualization artifacts of the raw-profit units (see text).\label{fig:gamma_profit}}
        {\includegraphics[width=.45\linewidth]{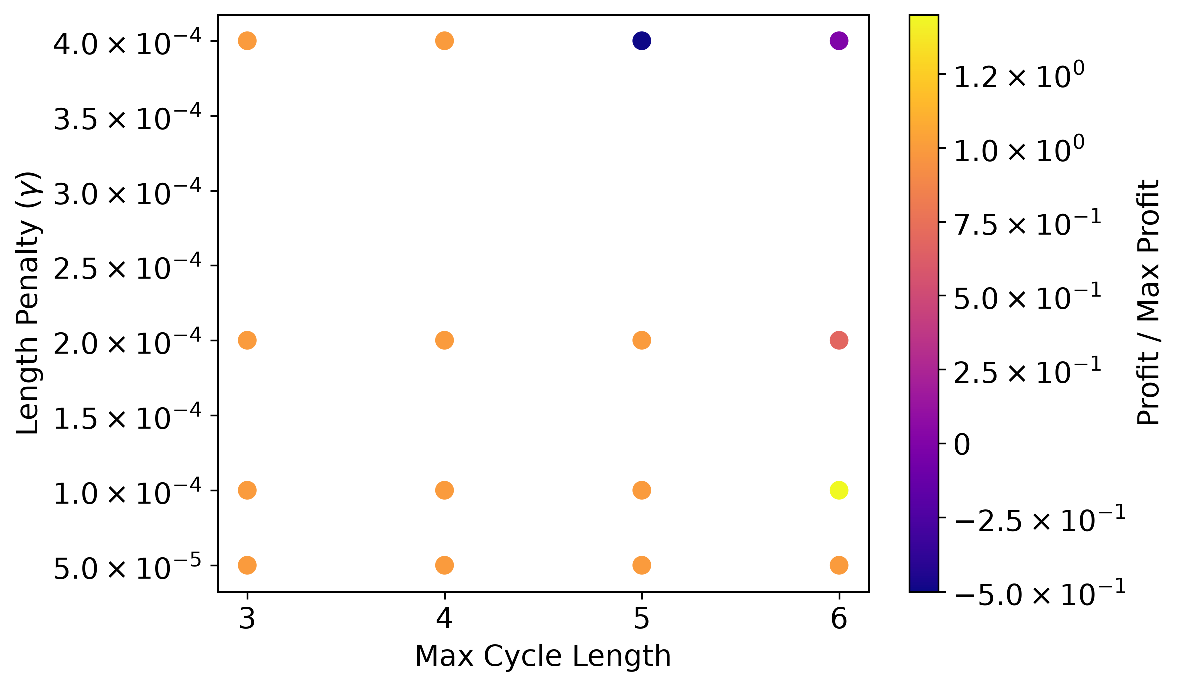}}        
    \end{tabular}
    \caption{Effect of the length-penalty coefficient $\gamma$ on the found cycle length, colored by the discrepancy from the optimal cycle length (a) and by the ratio of found to maximum raw profit (b).}
\end{figure}

\begin{figure}
    \centering
    \includegraphics[width=0.5\linewidth]{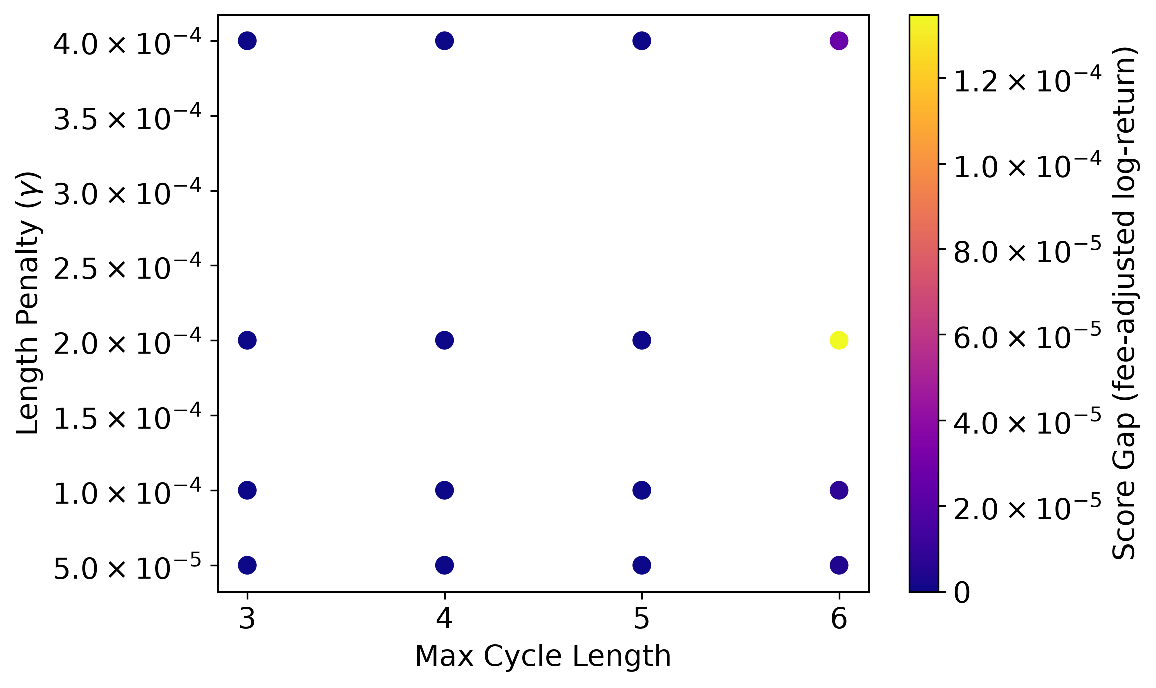}
    \caption{Gap in the fee-adjusted score between simulated annealing and the exact solution as a function of the length-penalty coefficient $\gamma$ and the maximum cycle length. The score is the objective both solvers optimize, so the gap is nonnegative by construction.}
    \label{fig:gamma_scoregap}
\end{figure}

Results up to this point were also evaluated only with no length penalty ($\gamma=0$). We present in Fig.~\ref{fig:gamma_length} how the found cycle length changes as a function of the length-penalty coefficient $\gamma$, with both the annealer and the exact baseline optimizing the same fee-adjusted objective of Eq.~\eqref{eq:hpfeasible}. Some care is needed when visualizing these runs in raw-profit units. With $\gamma>0$, the score-optimal cycle is deliberately shorter, and has lower raw profit, than the raw-profit-optimal cycle. A sample that lands on a longer, score-suboptimal or score-tied cycle can therefore display a raw-profit ratio exceeding one, bounded by $e^{\gamma\,\Delta m}$ for a length difference $\Delta m$. Conversely, when fees are large enough that every multi-currency cycle loses money net of fees, the score-optimal action collapses to a single-hop round trip. Such a round trip has exactly zero raw profit under reciprocal rates, so the plotted ratio collapses to zero or to floating-point noise. Both artifacts are visible in Fig.~\ref{fig:gamma_profit}, and neither indicates a solver error. We therefore also report in Fig.~\ref{fig:gamma_scoregap} the gap in the fee-adjusted score itself, which is the quantity both solvers optimize and is nonnegative by construction. Across the full $(\gamma, K)$ sweep, the gap never exceeds $1.3\times10^{-4}$ and is at floating-point zero for most grid points.

\subsection{Empirical comparison with prior QUBO encodings}\label{sec:encoding_comparison}
Table~\ref{tab:scaling} compares encodings by variable count and connectivity. In this section, we compare them by what the same sampler actually recovers. We implemented each prior formulation directly from its source paper: the edge-based and node-based models of Rosenberg~\cite{rosenberg2016}, the latter with its optional cycles-only constraint, the edge model of Deshpande, Das, and Mueller~\cite{deshpande2025}, the fixed-length model of Roy et al.~\cite{roy2025}, and the variable-length model of Mazzei et al.~\cite{mazzei2025}. Each implementation was validated by exhaustively enumerating its full ground-state spectrum on small instances. In each case, the ground state was checked against an independently computed optimum of that model's own objective: the maximum-weight disjoint cycle union for the edge-based model, the best fixed-length cycle for the node-based and Roy models, the best closed walk by dynamic programming for the Mazzei model, and the maximum-weight partial assignment for the Deshpande model. The penalty values are left open in \cite{deshpande2025,mazzei2025} and set by the Verma--Lewis heuristic in \cite{roy2025}, so we use validated sufficient defaults throughout. All models were then sampled by the same CPU \texttt{SimulatedAnnealingSampler} at matched settings of 1280 reads and 160 sweeps per run on the ten-currency, $K=6$, noise-$10^{-4}$ benchmark instance with gauge-fixed inputs (Sec.~\ref{sec:gauge}). Each sample was decoded to the best simple cycle of at most $K$ currencies and scored against the Held--Karp optimum of the identical fee-adjusted objective. Fixed-length formulations were swept over lengths $3,\dots,K$, and anchored formulations were swept over anchors.

\begin{table}[!h]
    \centering
    \begin{tabular}{lccccc}
        \toprule
        Encoding & Variables $\downarrow$ & Runs $\downarrow$ & Score gap $\downarrow$ & Sampler time (s) $\downarrow$ & Per-run time (s) $\downarrow$ \\
        \midrule
        Ours $(N{-}1)(K{-}1)$ & \textbf{45} & $10$ anchors & \textbf{0 (exact)} & 2.62 & 0.26 \\
        Ours - Worst of 10 & \textbf{45} & \textbf{1} & $3.3\times10^{-5}$ & 0.26 & 0.26 \\
        Roy et al.~\cite{roy2025} & 70 & $4$ lengths & $3.8\times10^{-6}$ & 0.87 & 0.22 \\
        1QBit node-based~\cite{rosenberg2016} & 60 & $4$ lengths & $8.0\times10^{-6}$ & 0.73  & 0.18 \\
        Mazzei et al.~\cite{mazzei2025} & 70 & \textbf{1} & $1.4\times10^{-5}$ & \textbf{0.16} & \textbf{0.16} \\
        Deshpande et al.~\cite{deshpande2025} & 90 & \textbf{1} & $3.7\times10^{-5}$ & 0.53 & 0.53 \\
        1QBit edge-based~\cite{rosenberg2016} & 90 & \textbf{1} & $5.5\times10^{-5}$ & 0.74 & 0.74 \\
        \bottomrule
    \end{tabular}
    \caption{All encodings of Table~\ref{tab:scaling} sampled by the same CPU simulated annealer at matched per-run settings (1280 reads, 160 sweeps) on the ten-currency, $K=6$ benchmark instance with gauge-fixed inputs and validated penalty weights. The gap is measured in the fee-adjusted log-return against the exact optimum of $7.106\times10^{-4}$; sampler time is the total sampling wall time over all runs for that encoding.}
    \label{tab:encoding_comparison}
\end{table}

The results appear in Table~\ref{tab:encoding_comparison}. The proposed encoding is the only one to recover the exact optimal cycle (zero score gap) and does so with the fewest logical variables, realizing the counts of Table~\ref{tab:scaling} at $N=10$ and $K=6$. The worst of the solutions found by the proposed encoding from the 10 different starting anchors also reaches a more optimal solution than formulations from 1QBit edge-based and Deshpande et al. indicating it's robust across anchors. Among the prior formulations, that of Roy et al.\ comes closest. This is consistent with it being the most tightly constrained, as it is the only prior model that enforces simple cycles outright. The Mazzei et al.\ formulation stands out on a different axis. It encodes all cycle lengths in a single QUBO by rewarding the repetition of a currency in consecutive positions, so it requires only one sampler run per instance and has the lowest total sampling time in the table. That repeat reward is exactly the per-transaction fee of Sec.~\ref{sec:qubo} with $\gamma/\alpha$ equal to the reward magnitude, so the two mechanisms for variable-length search are formally equivalent.

Implementing the prior models faithfully also surfaced two structural properties relevant to their interpretation. First, the Mazzei et al.\ Hamiltonian contains no constraint against revisiting a currency at non-adjacent positions, so its ground states are closed walks rather than simple cycles. On profit-rich instances, the minimizer laps a profitable sub-cycle repeatedly, and such samples must be discarded when the task is a simple cycle. Second, the Deshpande et al.\ Hamiltonian penalizes only out-degrees and in-degrees exceeding one and contains no flow-conservation term. Open paths are therefore zero-penalty configurations, and its exact ground state is the maximum-weight partial assignment rather than a cycle. Closed cycles appear only as components of that assignment and must be extracted in postprocessing, which is consistent with the critique in \cite{roy2025}. Both models, along with that of Roy et al., are descendants of the node-based closed-walk formulation of \cite{rosenberg2016}, which already noted the cycles-only constraint as an optional extension. Finally, gauge fixing benefits every encoding, not only ours. Without it, the same sampler budget leaves the single-run $O(N^2)$ and $O(NK)$ models with gaps in the $2$--$3\times10^{-4}$ range in our runs, roughly an order of magnitude worse than in Table~\ref{tab:encoding_comparison}. We caution that all entries derive from a single instance and sampler seed, and that the sampler's random stream depends on its thread count. Gaps at the $10^{-5}$ scale and below therefore fluctuate between runs and machines, and per-encoding success-probability statistics are a natural extension for the hardware study proposed below.

\section{Discussion and Future Work}
Finding CA solutions is of direct interest in finance \cite{schrimpf2019, chaboud2023, debelle2011, cartea2019, oomen2017, foucault2017, kozhan2012}. Simulated annealing is an approximate solution but can be used to evaluate the general efficacy of QUBO solutions for CA. We introduce a new solution which improves on existing work by placing the realistic constraint of beginning and ending cycles at a fixed reserve currency and by allowing for per-transaction trading-fee length penalties. We present this novel solution in Sec.~\ref{sec:qubo}. We also derive provably sufficient penalty weights for its constraint terms in Sec.~\ref{sec:weights} and introduce in Sec.~\ref{sec:gauge} an anchor-gauge reweighting that reduces the QUBO coefficient range to the arbitrage scale without changing the optimal cycle. The latter applies as preprocessing to any cycle-based CA encoding.

In Sec.~\ref{sec:qubo}, we also compare the scaling behavior in terms of number of fully connected variables and maximal connectivity among variables, showing that this formulation is competitive in terms of resource requirements. We also examine how this formulation embeds onto hardware quantum-annealing architectures such as the D-Wave Advantage Pegasus-16 graph. These embedding results suggest that hardware tests are feasible for problem sizes that are already challenging for exact enumeration, although actual time-to-solution would require hardware experiments including embedding overhead, chain-strength optimization, readout time, and success-probability analysis.

In Sec.~\ref{sec:sim}, we show a set of numerical evaluations on simulated annealing compared to the exact solution benchmark. We show in Fig.~\ref{fig:profit_noise} that simulated annealing is accurate even at the scale of a tenth of a penny of profit per dollar per cycle. We then show in Fig.~\ref{fig:profit_currencies_long} that, for constrained cycle lengths, the exact optimum was recovered by simulated annealing at 10240 reads for up to thirteen currencies, with a residual miss at fourteen. However, in Fig.~\ref{fig:time_currencies_long}, we show that the exact Held--Karp baseline is still two to three orders of magnitude faster in such cases.

Against the Held--Karp baseline on common hardware, classical simulated annealing does not outperform exact solution anywhere in the tested range. Simulated annealing does, however, generally find profitable solutions with the same order of magnitude of profit as the optimal solution. The asymptotics also identify the regime in which sampling becomes the only option. When the maximum cycle length grows with the number of currencies, exact search costs $O(2^{N-1}N^2)$ time and exponential memory per anchor, placing the classical wall near 20 to 25 currencies, while the anchored QUBO grows only polynomially. We further note that individual points in the profit figures derive from single sampler seeds and that the sampler's random stream depends on its thread count. A definitive comparison in the sampling regime therefore requires success-probability and time-to-solution statistics over instances and seeds. Because hardware quantum annealers can sample from different physical dynamics than classical simulated annealing, and because prior benchmarking has shown cases where quantum annealing can compare favorably with simulated annealing on related combinatorial optimization problems \cite{vert2024}, future work should evaluate this QUBO formulation directly on quantum-annealing hardware and compare time-to-solution, success probability, chain breaks, and solution quality against strong classical baselines.

To evaluate the behavior of the length-penalty term, we explore the effect of modifying the length penalty/trading fee coefficient $\gamma$. The behavior is as expected in Fig.~\ref{fig:gamma_length}: the found cycle length shortens as the penalty grows, with a consistent relationship between length and $\gamma$, and at fees exceeding all available per-edge profits the fee-adjusted optimum correctly collapses toward the shortest cycles. Measured in the objective both solvers optimize (Fig.~\ref{fig:gamma_scoregap}), the annealer tracks the exact fee-adjusted optimum to within $1.3\times10^{-4}$ across the entire sweep and exactly at most grid points, indicating that this parameter is not overly sensitive.

This problem and representation also have close ties to other optimal cycle-finding problems such as variations of the traveling salesman problem \cite{lucas2014}. Extensions of this work could include exploring minimal modifications needed to adapt the model to computationally expensive offline optimization problems. In particular, we would like to propose the task of finding maximally profitable shipping routes that would require minimal changes to account for fuel usage and costs.

\section{Conclusion}
CA using QUBO representations has been previously explored in the literature \cite{roy2025, rosenberg2016, mazzei2025, deshpande2025}. We build on these previous works by introducing realistic market constraints of anchoring the currencies to a fixed start currency and accounting for fixed trading fees. We show that this representation requires fewer logical qubits than previous solutions in the literature, with a maximal connectivity that is competitive with, and for short maximum cycle lengths lower than, that of other $O(NK)$ encodings. We further derive provably sufficient penalty weights for the constraint terms, verified by exhaustive ground-state enumeration. We also introduce an anchor-gauge reweighting that compresses the QUBO coefficient dynamic range by roughly three orders of magnitude on our benchmark instance while leaving the optimal cycle invariant. This directly addresses the analog-precision constraint of annealing hardware.

We present findings that the algorithm can successfully identify profitable cycles using simulated annealing, recovering the maximally profitable cycles at 10240 reads in all tested instances up to thirteen currencies at a maximum cycle length of six. In a matched-budget comparison against faithful, individually validated implementations of the five prior QUBO encodings of Table~\ref{tab:scaling}, the proposed encoding was the only one to recover the exact fee-adjusted optimum, and it did so at the lowest logical-variable count. We also show that the trading fee length penalty term behaves as expected in constraining the maximal sequence length when the penalty outweighs potential profit.

We propose that future work explore the use of hardware quantum annealing for this task. Specifically, our embedding analysis indicates that systems with 17 currencies and a maximal cycle length of 14 can fit on the D-Wave Advantage Pegasus-16 architecture. A naive exact-enumeration baseline for this problem would require checking over 59 trillion permutations, although the Held--Karp dynamic program used here still solves it in under a minute on a single CPU core, so the honest classical wall lies somewhat further out, near 20--25 currencies at unbounded cycle length. Demonstrating a practical quantum advantage, therefore, would require direct hardware runs and comparison against such optimized classical baselines using time-to-solution and success-probability metrics.

\section*{Code Availability}
The full code is made available at \href{https://github.com/ereinha/Forex-QUBO/tree/main}{ereinha/Forex-QUBO}.

\section*{AI Use Statement}
Claude Opus 4.8 was used for simulated review/criticism, spell-check, and grammar editing. Claude Opus 4.6-5.0 was used as a coding assistant for debugging, final code refactoring and plot formatting.

\section*{Acknowledgments}
The authors would like to thank Prof. Konstantin Matchev whose course on Quantum Computing at the University of Alabama inspired this work.

\bibliographystyle{unsrt}  
\bibliography{references}  

\appendix
\section{Data tables for the figures}\label{app:tables}
This appendix tabulates the quantities plotted in the figures of Sec.~\ref{sec:sim}, using the same numeric conventions as the figure axes.
\input{tables/tab_fig4_qubits}
\input{tables/tab_fig5_qubits_length}
\input{tables/tab_fig6a_steps}
\input{tables/tab_fig6b_shots}
\input{tables/tab_fig7_noise}
\input{tables/tab_fig8_profit_shots1280}
\input{tables/tab_fig8_profit_shots10240}
\input{tables/tab_fig9_walltime_shots1280}
\input{tables/tab_fig9_walltime_shots10240}
\input{tables/tab_fig10_length}
\input{tables/tab_fig11_12_gamma}

\end{document}

%% file: tables/tab_fig4_qubits.tex
\begin{table}[htbp]
    \centering
    \begin{tabular}{ccc}
        \toprule
        Currencies & Logical qubits & Physical qubits \\
        \midrule
        6 & 25 & 77 \\
        7 & 30 & 104 \\
        8 & 35 & 147 \\
        9 & 40 & 177 \\
        10 & 45 & 230 \\
        11 & 50 & 323 \\
        12 & 55 & 374 \\
        13 & 60 & 433 \\
        14 & 65 & 486 \\
        \bottomrule
    \end{tabular}
    \caption{Data plotted in Figs.~\ref{fig:fc_currencies} and~\ref{fig:pg16_currencies}: logical and Pegasus-16-embedded physical qubit counts vs number of currencies at a maximum cycle length of six.}
    \label{tab:data_qubits_currencies}
\end{table}

%% file: tables/tab_fig5_qubits_length.tex
\begin{table}[htbp]
    \centering
    \begin{tabular}{cccc}
        \toprule
        Max length & Logical qubits & Physical qubits & Max chain \\
        \midrule
        3 & 26 & 97 & 5 \\
        4 & 39 & 192 & 7 \\
        5 & 52 & 286 & 7 \\
        6 & 65 & 461 & 11 \\
        7 & 78 & 631 & 12 \\
        8 & 91 & 806 & 14 \\
        9 & 104 & 1149 & 17 \\
        10 & 117 & 1873 & 28 \\
        11 & 130 & 2499 & 37 \\
        12 & 143 & 2923 & 35 \\
        13 & 156 & 2940 & 32 \\
        14 & 169 & 3897 & 41 \\
        \bottomrule
    \end{tabular}
    \caption{Data plotted in Figs.~\ref{fig:fc_length} and~\ref{fig:pg16_length}: logical and Pegasus-16-embedded physical qubit counts vs maximum cycle length for 14 currencies, with the maximum embedding chain length.}
    \label{tab:data_qubits_length}
\end{table}

%% file: tables/tab_fig6a_steps.tex
\begin{table}[htbp]
    \centering
    \begin{tabular}{ccc}
        \toprule
        Sweeps & Exact profit & Simulated-anneal profit \\
        \midrule
        10 & $7.11\times10^{-4}$ & $6.74\times10^{-4}$ \\
        20 & $7.11\times10^{-4}$ & $7.11\times10^{-4}$ \\
        40 & $7.11\times10^{-4}$ & $7.07\times10^{-4}$ \\
        80 & $7.11\times10^{-4}$ & $7.03\times10^{-4}$ \\
        160 & $7.11\times10^{-4}$ & $7.11\times10^{-4}$ \\
        320 & $7.11\times10^{-4}$ & $7.11\times10^{-4}$ \\
        640 & $7.11\times10^{-4}$ & $7.11\times10^{-4}$ \\
        1280 & $7.11\times10^{-4}$ & $7.07\times10^{-4}$ \\
        \bottomrule
    \end{tabular}
    \caption{Data plotted in Fig.~\ref{fig:profit_vs_steps}: profit vs number of simulated-anneal sweeps for ten currencies and a maximum cycle length of six.}
    \label{tab:data_steps}
\end{table}

%% file: tables/tab_fig6b_shots.tex
\begin{table}[htbp]
    \centering
    \begin{tabular}{ccc}
        \toprule
        Reads & Exact profit & Simulated-anneal profit \\
        \midrule
        10 & $7.11\times10^{-4}$ & $4.06\times10^{-4}$ \\
        20 & $7.11\times10^{-4}$ & $4.46\times10^{-4}$ \\
        40 & $7.11\times10^{-4}$ & $6.74\times10^{-4}$ \\
        80 & $7.11\times10^{-4}$ & $6.74\times10^{-4}$ \\
        160 & $7.11\times10^{-4}$ & $6.74\times10^{-4}$ \\
        320 & $7.11\times10^{-4}$ & $6.74\times10^{-4}$ \\
        640 & $7.11\times10^{-4}$ & $6.74\times10^{-4}$ \\
        1280 & $7.11\times10^{-4}$ & $7.11\times10^{-4}$ \\
        2560 & $7.11\times10^{-4}$ & $7.11\times10^{-4}$ \\
        \bottomrule
    \end{tabular}
    \caption{Data plotted in Fig.~\ref{fig:profit_vs_shots}: profit vs number of reads for ten currencies and a maximum cycle length of six.}
    \label{tab:data_shots}
\end{table}

%% file: tables/tab_fig7_noise.tex
\begin{table}[htbp]
    \centering
    \begin{tabular}{ccc}
        \toprule
        Noise & Exact profit & Simulated-anneal profit \\
        \midrule
        $1.00\times10^{-5}$ & $7.11\times10^{-5}$ & $7.11\times10^{-5}$ \\
        $1.00\times10^{-4}$ & $7.11\times10^{-4}$ & $7.11\times10^{-4}$ \\
        $1.00\times10^{-3}$ & $7.13\times10^{-3}$ & $7.05\times10^{-3}$ \\
        $1.00\times10^{-2}$ & $7.37\times10^{-2}$ & $7.32\times10^{-2}$ \\
        $1.00\times10^{-1}$ & $1.04\times10^{0}$ & $1.04\times10^{0}$ \\
        \bottomrule
    \end{tabular}
    \caption{Data plotted in Fig.~\ref{fig:profit_noise}: profit vs monetary noise for ten currencies and a maximum cycle length of six.}
    \label{tab:data_noise}
\end{table}

%% file: tables/tab_fig8_profit_shots1280.tex
\begin{table}[htbp]
    \centering
    \begin{tabular}{ccc}
        \toprule
        Currencies & Exact profit & Simulated-anneal profit \\
        \midrule
        6 & $5.80\times10^{-4}$ & $5.80\times10^{-4}$ \\
        7 & $6.72\times10^{-4}$ & $6.72\times10^{-4}$ \\
        8 & $5.95\times10^{-4}$ & $5.95\times10^{-4}$ \\
        9 & $7.93\times10^{-4}$ & $7.93\times10^{-4}$ \\
        10 & $7.11\times10^{-4}$ & $7.11\times10^{-4}$ \\
        11 & $8.60\times10^{-4}$ & $8.60\times10^{-4}$ \\
        12 & $8.11\times10^{-4}$ & $7.46\times10^{-4}$ \\
        13 & $8.79\times10^{-4}$ & $8.65\times10^{-4}$ \\
        14 & $8.99\times10^{-4}$ & $7.99\times10^{-4}$ \\
        \bottomrule
    \end{tabular}
    \caption{Data plotted in Fig.~\ref{fig:profit_currencies_short}: profit vs number of currencies with 1280 reads.}
    \label{tab:data_profit_currencies_1280}
\end{table}

%% file: tables/tab_fig8_profit_shots10240.tex
\begin{table}[htbp]
    \centering
    \begin{tabular}{ccc}
        \toprule
        Currencies & Exact profit & Simulated-anneal profit \\
        \midrule
        6 & $5.80\times10^{-4}$ & $5.80\times10^{-4}$ \\
        7 & $6.72\times10^{-4}$ & $6.72\times10^{-4}$ \\
        8 & $5.95\times10^{-4}$ & $5.95\times10^{-4}$ \\
        9 & $7.93\times10^{-4}$ & $7.93\times10^{-4}$ \\
        10 & $7.11\times10^{-4}$ & $7.11\times10^{-4}$ \\
        11 & $8.60\times10^{-4}$ & $8.60\times10^{-4}$ \\
        12 & $8.11\times10^{-4}$ & $8.11\times10^{-4}$ \\
        13 & $8.79\times10^{-4}$ & $8.79\times10^{-4}$ \\
        14 & $8.99\times10^{-4}$ & $8.99\times10^{-4}$ \\
        \bottomrule
    \end{tabular}
    \caption{Data plotted in Fig.~\ref{fig:profit_currencies_long}: profit vs number of currencies with 10240 reads.}
    \label{tab:data_profit_currencies_10240}
\end{table}

%% file: tables/tab_fig9_walltime_shots1280.tex
\begin{table}[htbp]
    \centering
    \begin{tabular}{ccc}
        \toprule
        Currencies & Exact (s) & Simulated anneal (s) \\
        \midrule
        6 & 0.01 & 0.66 \\
        7 & 0.02 & 0.92 \\
        8 & 0.04 & 1.27 \\
        9 & 0.08 & 1.73 \\
        10 & 0.14 & 2.27 \\
        11 & 0.26 & 2.90 \\
        12 & 0.41 & 3.48 \\
        13 & 0.67 & 4.22 \\
        14 & 1.03 & 5.22 \\
        \bottomrule
    \end{tabular}
    \caption{Data plotted in Fig.~\ref{fig:time_currencies_short}: wall time vs number of currencies with 1280 reads, on the same CPU.}
    \label{tab:data_walltime_currencies_1280}
\end{table}

%% file: tables/tab_fig9_walltime_shots10240.tex
\begin{table}[htbp]
    \centering
    \begin{tabular}{ccc}
        \toprule
        Currencies & Exact (s) & Simulated anneal (s) \\
        \midrule
        6 & 0.01 & 5.12 \\
        7 & 0.02 & 7.40 \\
        8 & 0.04 & 10.36 \\
        9 & 0.07 & 14.07 \\
        10 & 0.15 & 17.52 \\
        11 & 0.25 & 22.96 \\
        12 & 0.44 & 28.35 \\
        13 & 0.67 & 34.76 \\
        14 & 1.05 & 40.55 \\
        \bottomrule
    \end{tabular}
    \caption{Data plotted in Fig.~\ref{fig:time_currencies_long}: wall time vs number of currencies with 10240 reads, on the same CPU.}
    \label{tab:data_walltime_currencies_10240}
\end{table}

%% file: tables/tab_fig10_length.tex
\begin{table}[htbp]
    \centering
    \begin{tabular}{ccccc}
        \toprule
        Max length & Exact (s) & Simulated anneal (s) & Exact profit & Simulated-anneal profit \\
        \midrule
        3 & 0.09 & 14.00 & $4.06\times10^{-4}$ & $4.06\times10^{-4}$ \\
        4 & 0.19 & 22.84 & $6.11\times10^{-4}$ & $6.11\times10^{-4}$ \\
        5 & 0.48 & 31.46 & $8.70\times10^{-4}$ & $8.70\times10^{-4}$ \\
        6 & 1.01 & 40.31 & $8.99\times10^{-4}$ & $8.99\times10^{-4}$ \\
        7 & 1.97 & 48.41 & $9.82\times10^{-4}$ & $8.99\times10^{-4}$ \\
        8 & 3.11 & 56.27 & $1.07\times10^{-3}$ & $9.52\times10^{-4}$ \\
        9 & 4.32 & 65.90 & $1.32\times10^{-3}$ & $1.22\times10^{-3}$ \\
        10 & 4.98 & 73.20 & $1.52\times10^{-3}$ & $1.31\times10^{-3}$ \\
        11 & 5.77 & 82.73 & $1.56\times10^{-3}$ & $1.27\times10^{-3}$ \\
        \bottomrule
    \end{tabular}
    \caption{Data plotted in Figs.~\ref{fig:time_length} and~\ref{fig:profit_length}: wall time and profit vs maximum cycle length for 14 currencies with 10240 reads, on the same CPU.}
    \label{tab:data_length}
\end{table}

%% file: tables/tab_fig11_12_gamma.tex
\begin{table}[htbp]
    \centering
    \begin{tabular}{ccccccc}
        \toprule
        Max length & $\gamma$ & Optimal length & Found length & Exact profit & Simulated-anneal profit & Score gap \\
        \midrule
        3 & $5.00\times10^{-5}$ & 3 & 3 & $4.62\times10^{-4}$ & $4.62\times10^{-4}$ & $8.88\times10^{-16}$ \\
        3 & $1.00\times10^{-4}$ & 3 & 3 & $4.62\times10^{-4}$ & $4.62\times10^{-4}$ & $8.88\times10^{-16}$ \\
        3 & $2.00\times10^{-4}$ & 3 & 3 & $4.62\times10^{-4}$ & $4.62\times10^{-4}$ & $8.88\times10^{-16}$ \\
        3 & $4.00\times10^{-4}$ & 3 & 3 & $4.62\times10^{-4}$ & $4.62\times10^{-4}$ & $8.88\times10^{-16}$ \\
        4 & $5.00\times10^{-5}$ & 4 & 4 & $6.07\times10^{-4}$ & $6.07\times10^{-4}$ & $8.88\times10^{-16}$ \\
        4 & $1.00\times10^{-4}$ & 4 & 4 & $6.07\times10^{-4}$ & $6.07\times10^{-4}$ & $8.88\times10^{-16}$ \\
        4 & $2.00\times10^{-4}$ & 3 & 3 & $4.16\times10^{-4}$ & $4.16\times10^{-4}$ & $3.33\times10^{-16}$ \\
        4 & $4.00\times10^{-4}$ & 3 & 3 & $4.16\times10^{-4}$ & $4.16\times10^{-4}$ & $3.33\times10^{-16}$ \\
        5 & $5.00\times10^{-5}$ & 5 & 5 & $6.56\times10^{-4}$ & $6.56\times10^{-4}$ & $1.33\times10^{-15}$ \\
        5 & $1.00\times10^{-4}$ & 4 & 4 & $5.92\times10^{-4}$ & $5.92\times10^{-4}$ & $1.33\times10^{-15}$ \\
        5 & $2.00\times10^{-4}$ & 4 & 4 & $5.92\times10^{-4}$ & $5.92\times10^{-4}$ & $1.33\times10^{-15}$ \\
        5 & $4.00\times10^{-4}$ & 2 & 2 & $2.22\times10^{-16}$ & $-1.11\times10^{-16}$ & $1.78\times10^{-15}$ \\
        6 & $5.00\times10^{-5}$ & 6 & 6 & $7.11\times10^{-4}$ & $7.07\times10^{-4}$ & $3.78\times10^{-6}$ \\
        6 & $1.00\times10^{-4}$ & 3 & 3 & $4.27\times10^{-4}$ & $4.27\times10^{-4}$ & $5.55\times10^{-16}$ \\
        6 & $2.00\times10^{-4}$ & 3 & 3 & $4.27\times10^{-4}$ & $4.27\times10^{-4}$ & $5.55\times10^{-16}$ \\
        6 & $4.00\times10^{-4}$ & 3 & 2 & $4.27\times10^{-4}$ & 0 & $2.69\times10^{-5}$ \\
        \bottomrule
    \end{tabular}
    \caption{Data plotted in Figs.~\ref{fig:gamma_length}, \ref{fig:gamma_profit}, and~\ref{fig:gamma_scoregap}: found and optimal cycle length, profits, and fee-adjusted score gap for each length penalty $\gamma$ and maximum cycle length.}
    \label{tab:data_gamma}
\end{table}